\documentclass[aps,prb,amsmath,twocolumn,superscriptaddress,showpacs]{revtex4-1}
\usepackage{color}
\usepackage{amssymb}
\usepackage{dcolumn}
\usepackage{amstext}
\usepackage{graphicx}
\usepackage{amsmath}
\usepackage[colorlinks=true,urlcolor=blue,anchorcolor=blue,linkcolor=red,citecolor=red,breaklinks=true]{hyperref}
\usepackage{amsthm}
\usepackage{longtable}
\usepackage{amsfonts}
\usepackage{amssymb}
\usepackage{subfigure}
\usepackage{latexsym}
\usepackage{ulem}

\usepackage{physics}
\newcommand{\RN}[1]{%
  \textup{\uppercase\expandafter{\romannumeral#1}}%
}

\begin{document}

\title{Effects of spin orbit coupling on proximity induced superconductivity}

\author{Vivek Mishra}
\affiliation{Kavli Institute for Theoretical Sciences, University of Chinese Academy of Sciences, Beijing 100190, China}

\author{Yu Li}
\affiliation{Kavli Institute for Theoretical Sciences, University of Chinese Academy of Sciences, Beijing 100190, China}

\author{Fu-Chun Zhang}
\email{fuchun@ucas.ac.cn}
\affiliation{Kavli Institute for Theoretical Sciences, University of Chinese Academy of Sciences, Beijing 100190, China}
\affiliation{CAS Center for Excellence in Topological Quantum Computation, University of Chinese Academy of Sciences, Beijing 100190, China}

\author{Stefan Kirchner}
\email{stefan.kirchner@correlated-matter.com}
\affiliation{Department of Electrophysics \& Center for Theoretical and Computational Physics, National Yang Ming Chiao Tung University, Hsinchu 30010, Taiwan}
\affiliation{Center for Emergent Functional Matter Science, National Yang Ming Chiao Tung University, Hsinchu 30010, Taiwan}

\date{\today}

\begin{abstract}
We investigate the effect of spin orbit coupling on proximity induced superconductivity in a normal metal attached to a superconductor. Specifically, we consider a heterostructure where the presence of interfaces gives rise to a Rashba spin orbit coupling.  The properties of the induced superconductivity in these systems are addressed within the tunneling Hamiltonian formalism. We find that the spin orbit coupling induces a mixture of singlet and triplet pairing and, under specific circumstances, an odd frequency, even parity, spin triplet pairs can arise. We also address the effect of impurity scattering on the induced pairs, and discuss our results in context of heterostructures consisting of materials with  spin-momentum locking.
\end{abstract}

\date{\today}

\maketitle

\section{Introduction}
Hybrid nanostructures consisting of superconductors have been intensively studied, both experimentally and theoretically. Such hybrid proximity structures provide a platform  for the realization of novel superconducting states in the vicinity of interfaces that connect  superconductors to  non-superconducting materials. The current heightened interest in these systems is also driven by the potential of these heterostructures to host Majorana fermions\cite{PW1,PW2,PW3,PW4}. As these Majorana fermions obey non-Abelian braiding statistics they may serve as  building blocks for fault-tolerant quantum computation \cite{Wilczek2009,Kitaev2001,Ivanov2001,SCZ2011,Rodrigo2012}. Superconductor and ferromagnetic (S/F) hybrid structures have been studied heavily\cite{SF0,SF0a,Khaire2010,SF2,SF1,Buzdin2005}, with a recent focus on proximity structures with topological materials. In such proximity structures, the spin orbit coupling (SOC) plays an important role. It is usually induced by the breaking of inversion symmetry {\itshape e.g.} through an underlying substrate or the presence of an interface. In a superconductor, the SOC leads to a mixing of singlet and triplet pairing.
In this two-component superconductivity either singlet or triplet pairing can be dominating\cite{Bauer2004,Frigeri2004}.  Recent observation of triplet dominant two component superconductivity in  CoSi$_2$/TiSi$_2$/Si heterostructures has confirmed the realization of dominant triplet pairing in these structures via a  substrate induced SOC\cite{Chiu2021,Chiu2022}. 

The role of SOC in the S/F nanostructures has been studied extensively \cite{Tokatly2013,Tokatly2014,Jacobsen2015,Alidoust2015,Arjoranta2016}, where the junctions involved $s$-wave superconductors. These studies were carried out in the quasi-classical formalism in the diffusive limit\cite{Usadel1970}.
 In the diffusive limit, the most dominant energy scale in the problem is the elastic impurity scattering {\itshape i. e.} the  impurity scattering rate $\tau^{-1} \gg \Delta, \tau^{-1}_{sf}, \tau^{-1}_{in} $, where $\Delta$ is the superconducting gap, $\tau^{-1}_{sf/in}$ is the scattering rate from spin-flip/inelastic scattering. In such a regime, even frequency - spin singlet  - even parity superconductor (ESE) induces ESE  pairs in the diffusive normal metals, and even frequency  - spin triplet - odd parity superconductors (ETO) leads to the formation of odd frequency  - spin triplet - even parity (OTE) pairs in the diffusive metal,  as long as the interface is non-magnetic in nature\cite{Tanaka2007}. A non-magnetic interface prevents triplet to singlet conversion. Our main objective is to understand the properties of proximity-induced superconductivity in metals with sizable SOC. We will examine the stability of the proximity-induced superconductivity against weak disorder and analyze the emergence of odd frequency pairs. 
{The effect of SOC on the superconducting side has been  explored in Refs.\,\onlinecite{Tamura2019,Mishra2021} while interfaces with SOC have been studied within the BTK formalism \cite{Kapri2017,Samokhin2010}. These studies do not include the effect of proximity induced pairs. The effect of SOC in the normal metal side on proximity induced superconductivity has  so far nor been addressed. In general, both cases, {\itshape i.e.}, SOC only at the interface and SOC in the normal metal side need to be distinguished. Whereas the first leads to a spatially localized spin-active boundary condition, the second results in a reconstruction of the bands and broken spin rotational symmetry in the normal metal.}
 
 In this paper, we focus on the effect of SOC on induced superconductivity in a normal non-magnetic metal that is connected to an unconventional superconductor. For concreteness, we consider a Rashba SOC that is induced by the underlying substrate beneath the normal metal component. One reason for this particular choice of SOC interaction is  that it can be generated and controlled by applying a gate voltage to the heterostructure.   Here, we focus on the properties of the induced superconductivity in the normal metal connected to a triplet superconductor. In what follows, we adopt the tunneling Hamiltonian formalism\cite{McMillan1968,Stanescu2011,Balatsky2012,Yokoyama2012,Tkachov2013,YuWu2016,Hugdal2019}. In the next section, we introduce the basic model and the theoretical methods.  The subsequent section provides a  discussion of our results. The final section summarizes the key qualitative conclusions.

 \begin{figure}
\includegraphics[width=.88\linewidth]{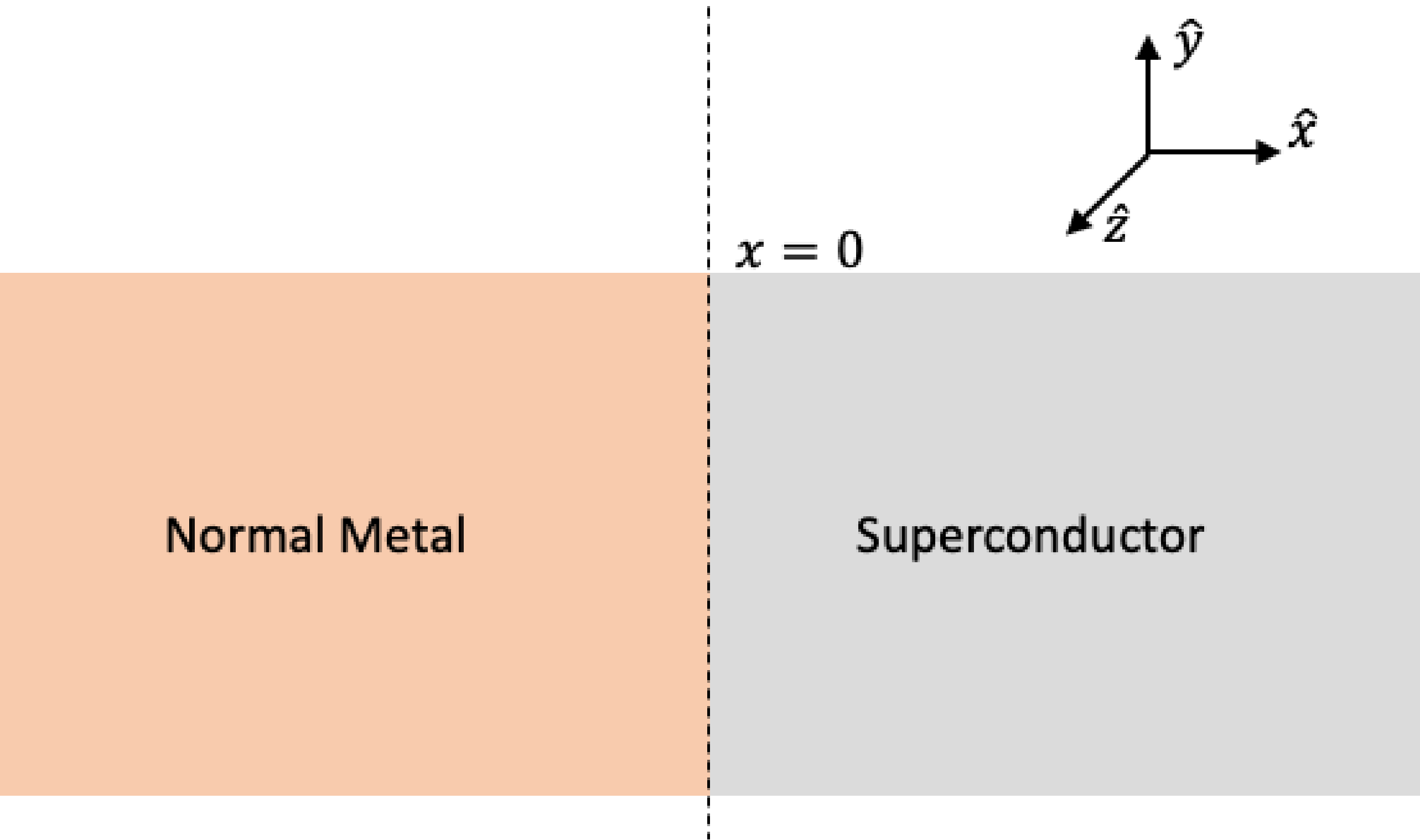}
\caption{Schematic illustration of a superconductor - normal junction. The interface is along the $yz$ plane.}
\label{Fig:SN_junc}
\end{figure}

\section{Model \& Formalism}\label{sec:MF}
Figure \ref{Fig:SN_junc} shows a schematic diagram of the superconductor - normal (SN) junction.  The interface is located at the $x=0$ plane. We consider a SOC that is  induced by the substrate and take  the  $\hat{\mathbf{z}}$ axis to be parallel the  substrate normal. The Hamiltonian for the normal component reads,
\begin{eqnarray}
{H}_N &=& \sum_{\mathbf{k},\sigma} c^\dagger_{k\sigma} \left[ \xi_\mathbf{k} \delta_{\sigma \sigma'}+ \hat{H} _{Rashba} \right] c_{k\sigma'},
\end{eqnarray}
where $c^\dagger/c$ is the electron creation/annihilation operator, $\xi_\mathbf{k} $ is the electronic dispersion, $\mathbf{k}$ denotes the momentum and $\sigma$ represents the electron spin. We denote a $4\times4$ matrix in Nambu-spin space with $\check{\square}$ while $\hat{\square}$ indicates a $2\times2$ matrix in the spin space. The Rashba SOC term reads,
\begin{eqnarray}
\hat{\mathcal{H}}_{Rashba} &=& -\frac{\alpha}{m} \left( \boldsymbol{\sigma} \times \mathbf{k}\right)\cdot \hat{\mathbf{z}} = \epsilon_N  \left(  \hat{\mathbf{z}} \times \hat{\mathbf{k}}\right)\cdot \boldsymbol{\sigma}  , \\
\epsilon_N&=&\frac{\alpha |\mathbf{k}|}{m}.
\end{eqnarray} 
Here, $\alpha$ is the Rashba SOC coupling constant, $m$ is the effective mass and $\boldsymbol{\sigma}$ is the Pauli vector ($\sigma_x,\sigma_y,\sigma_z$), where $\sigma_{x/y/z}$ are the Pauli matrices in spin space. The  two helical bands generated by this term have energies  $\xi_\mathbf{k}\pm \epsilon_N$.  The Hamiltonian for the superconductor reads,
\begin{eqnarray}
{H}_{sc} &=& \Psi^\dagger \begin{pmatrix}
\xi_{\mathbf{k}} \sigma_0 & \hat{\Delta} \\
\hat{\Delta}^\dagger & -\xi_{\mathbf{k}} \sigma_0
\end{pmatrix} \Psi.
\label{eq:Hsc}
\end{eqnarray}
Here, $\sigma_0$ is the $2\times2$ identity matrix  in  spin space,  $\xi_{\mathbf{k}}$ is the  dispersion in the superconductor, $\Psi^\dagger=(a^\dagger_{\uparrow \mathbf{k}},a^\dagger_{\downarrow \mathbf{k}},a_{\uparrow -\mathbf{k}},a_{\downarrow -\mathbf{k}})$, where $a^\dagger/a$ is the creation/annihilation operator. The gap $\hat{\Delta}=\Delta i\sigma_y$ for the singlet case while for the triplet case $\hat{\Delta}=\Delta \mathbf{d}\cdot \boldsymbol{\sigma} i \sigma_y$. Here, $\mathbf{d}$ is the order-parameter vector in  spin space of the triplet paring.  The tunneling Hamiltonian is,
\begin{equation}
{H}_{tunneling} = \gamma \Psi^\dagger \check{\tau}_3 \Phi + h. c.
\label{eq:tunn}
\end{equation}
where  {$\check{\tau}_3=\mathrm{diag}(1,1,-1,-1)$}  is a  matrix in  Nambu spin space, $\Phi^\dagger =(c^\dagger_{\uparrow \mathbf{k}},c^\dagger_{\downarrow \mathbf{k}},c_{\uparrow -\mathbf{k}},c_{\downarrow -\mathbf{k}})$,  where $c^\dagger_{\mathbf{k}\sigma}$ and $c_{\mathbf{k}\sigma}$ are the creation and annihilation operators in the normal metal segment and $\gamma$ is the tunneling matrix element. We assume $\gamma$ to be spin and momentum independent and take  it to be real. 

The mean-field expression for the Green's function of the superconducting component is
\begin{equation}
\check{G}_{sc} =  \frac{1}{\omega^2-\xi^2-|\Delta|^2} \begin{pmatrix}
(\omega+\xi) \sigma_0 & \hat{\Delta} \\
\hat{\Delta}^\dagger & (\omega-\xi) \sigma_0
\end{pmatrix}.
\label{Eq:Gsc}
\end{equation}
For notational convenience, we abbreviate $\xi_{\mathbf{k}}$ as $\xi$. The tunneling self energy for the normal side of the junction at the interface reads,
\begin{eqnarray}
\check{\Sigma}_t(k_{||},\omega) &=& |\gamma|^2 \int \frac{d k_\perp}{2 \pi} \frac{1}{\omega^2-\xi^2-|\Delta|^2} \nonumber \\
&\times&  \begin{pmatrix}
(\omega+\xi) \sigma_0 & -\hat{\Delta} \\
-\hat{\Delta}^\dagger & (\omega-\xi) \sigma_0
\end{pmatrix}.
\label{eq:Self_tunn}
\end{eqnarray}
Here, $k_\perp$ is the momentum component perpendicular to the SN interface and $k_{||}$ is the momentum parallel to it.
The self-energy has the same general structure  as the structure of the Green's function in the superconductor. The integration over the momentum component perpendicular to the interface may modify $\mathbf{d}$ to $\tilde{\mathbf{d}}$ depending on the junction geometry. 
 
To study the robustness of the induced pairs against disorder, we consider point-like impurities randomly distributed in the normal metal. Within a self-consistent T-matrix approximation which incorporates all scattering processes from a single impurity site the impurity self energy contribution is,
\begin{eqnarray}
\check{\Sigma}_{imp}( \omega) &=& n_{imp} \check{\tau}_3 {V}_{imp}\left[ \check{1}- \check{\mathbf{g}} \check{\tau}_3 {V}_{imp} \right]^{-1},
\end{eqnarray}
where $n_{imp}$ is the impurity concentration, $V_{imp}$ is the impurity potential, $\check{1}$ is the $4\times4$ identity matrix  and $\check{\mathbf{g}}$ is
\begin{eqnarray}
    \check{\mathbf{g}} &=& \int_{\mathbf{k}} \check{\mathbb{G}}.
\end{eqnarray}
where
\begin{eqnarray}
\check{\mathbb{G}}&=& \frac{\check{1}}{\check{G}_0^{-1}-\check{\Sigma}_t( k_{||},\omega)-\check{\Sigma}_{imp}(\omega)}.
\end{eqnarray}
Here $\check{G}_0$ is the normal metal bare Green's function, and the impurity self-energy is calculated self-consistently. { The order-parameter energy scale $\Delta$ sets the energy scale in this problem, and not calculated self-consistently. It is assumed that the superconductor has an effective attractive interaction to generate the gap symmetry under consideration. A self-consistent determination of $\Delta$ will only change its numerical value. A change in the spin-momentum structure of the order parameter is not expected  as there is no attractive interaction that could generate any changes in the order parameter structure.}

\section{Results \& Discussion}
\subsection{SN junction with a singlet superconductor}\label{sec:R1}
First, we consider a pure singlet superconductor attached to the normal metal component discussed above. In that case,
$\hat{\Delta}=\Delta i \sigma_y$ and  the tunneling self-energy is given by,
\begin{eqnarray}
\check{\Sigma}_t(k_{||},\omega) = |\gamma|^2 \begin{pmatrix}
(\Sigma_0 + \Sigma_3)\sigma_0 & i \sigma_y {\Sigma}_1 \\
-i \sigma_y {\Sigma}_1 & (\Sigma_0 - \Sigma_3)\sigma_0. 
\end{pmatrix}
\end{eqnarray}
In the case of
a momentum independent $\Delta$ and for a particle-hole symmetric system, the tunneling self-energy becomes,
\begin{eqnarray}
\check{\Sigma}_t(k_{||},\omega) = -\Gamma_t \frac{1}{\sqrt{\Delta^2-\omega^2}} \begin{pmatrix}
\omega \sigma_0 & i \sigma_y {\Delta} \\
-i \sigma_y \Delta &\omega \sigma_0 
\end{pmatrix}.
\label{eq:Sigma_singlet}
\end{eqnarray}
Here $\Gamma_t\equiv \pi |\gamma|^2 \nu_s$ is the energy scale associated with the tunneling process, where $\nu_s$ is the normal state density of states (DOS) of the superconductor at the Fermi level.  Inclusion of particle-hole asymmetry gives finite $\Sigma_3$, that can be absorbed in the chemical potential. As it turns out, the presence of particle-hole asymmetry  does not lead to significant qualitative difference.  The normal metal Green's function in the clean limit is,
\begin{eqnarray}
\check{\mathbb{G}}  &=& \begin{pmatrix}
\left[ \bar{\omega} - {\xi} \right] \sigma_0 - \epsilon_N (\mathbf{w}  \cdot \boldsymbol{\sigma})   & - i\Sigma_1 \sigma_y \\
i \Sigma_1 \sigma_y & \left[\bar{\omega} + {\xi} \right] \sigma_0 - \epsilon_N (\mathbf{w}  \cdot \boldsymbol{\sigma}^\ast) 
\end{pmatrix}^{-1}, \nonumber \\
\check{\mathbb{G}} &=& \begin{pmatrix}
\hat{\mathbb{G}}_{11} & \hat{\mathbb{G}}_{12} \\
\hat{\mathbb{G}}_{21} & \hat{\mathbb{G}}_{22}.
\end{pmatrix}
\end{eqnarray}
Here, $\bar{\omega}=\omega-\Sigma_0$ and $\mathbf{w}=\hat{\mathbf{z}}\times \hat{\mathbf{k}}$.  The components of $\check{\mathbb{G}} $ are,
\begin{eqnarray}
\hat{\mathbb{G}}_{11}  &=& \frac{1}{2} \left[ \mathcal{R}_+ + \mathcal{R}_- \right] +  \frac{1}{2} \left[ \mathcal{R}_+ - \mathcal{R}_- \right] (\mathbf{w}  \cdot \boldsymbol{\sigma}), \label{Eq:GN_sing_tunn0} \\
\mathcal{R}_\pm &=& \frac{\bar{\omega} + (\bar{\xi} \pm \epsilon_N)}{\bar{\omega}^2-(\bar{\xi} \pm \epsilon_N)^2 -\Sigma_1^2}, \\
\hat{\mathbb{G}}_{12} & =&  \left (\frac{1}{2} \left[ \frac{1}{D_+} + \frac{1}{D_-}\right] + \frac{1}{2} \left[ \frac{1}{D_+} - \frac{1}{D_-}\right] \mathbf{w} \cdot   \boldsymbol{\sigma} \right), \nonumber \\
& \times& ( i \sigma_y  \Sigma_1)\label{Eq:GN_sing_tunn1}  \\
D_\pm &=& \bar{\omega}^2-(\bar{\xi} \pm \epsilon_N)^2 -\Sigma_1^2. 
\end{eqnarray}
The structure of the induced pairs can be obtained from the anomalous Green's function $\hat{\mathbb{G}}_{12} $. The first term of $\hat{\mathbb{G}}_{12}$ in Eq. \eqref{Eq:GN_sing_tunn1} is the spin-singlet, even parity and even frequency component. This is the conventional proximity effect for a singlet superconductor. The second term of $\hat{\mathbb{G}}_{12} $ is a spin-triplet, odd parity and even frequency term. This term is directly proportional  to the strength of SOC. Due to finite SOC, spin rotational symmetry is broken which allows for a mixing  of singlet and triplet terms. The $\mathbf{d}$-vector for triplet pairing is determined by the SOC vector $\mathbf{w}$.

\begin{figure}
\includegraphics[width=.95\linewidth]{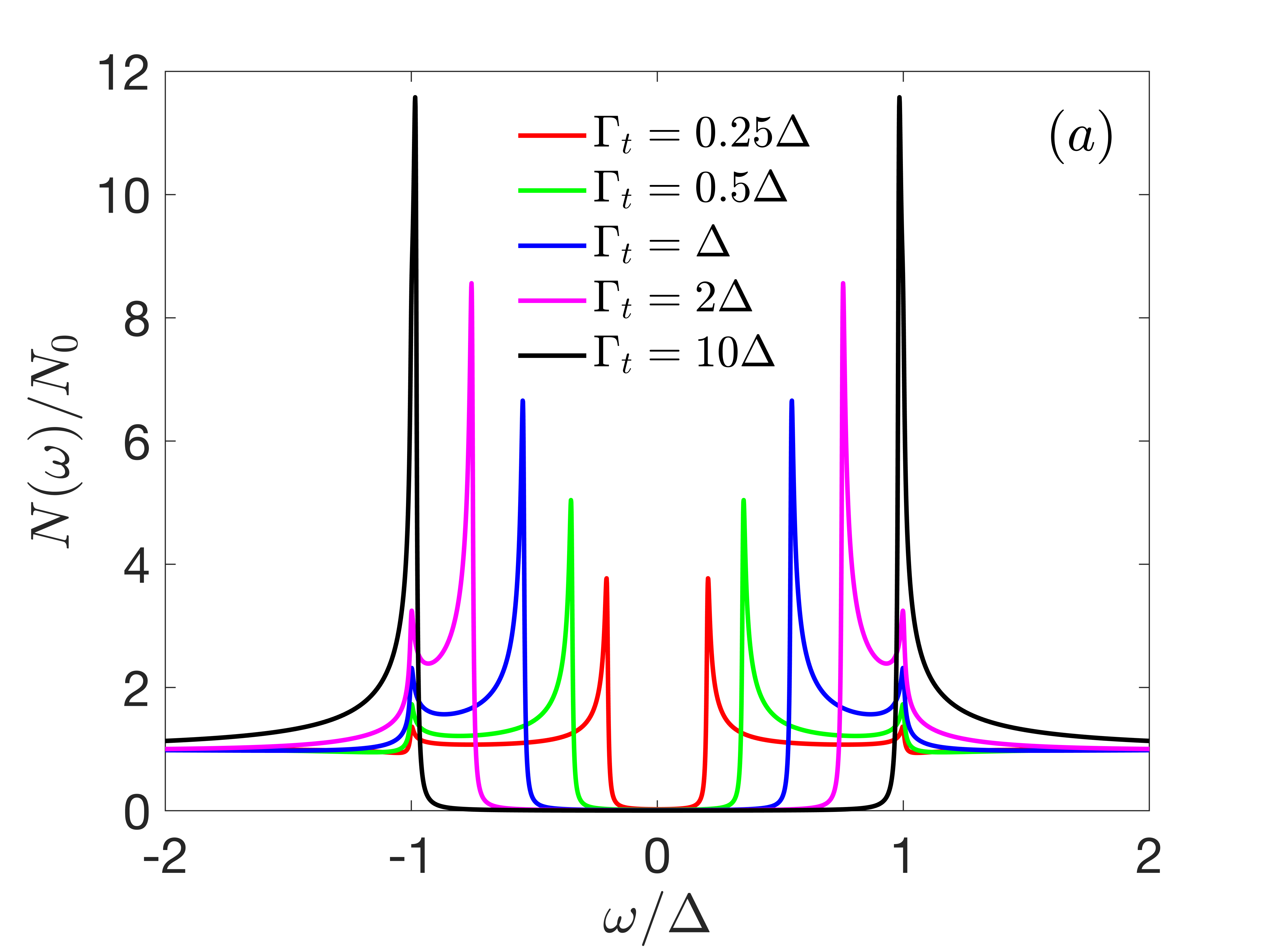}
\includegraphics[width=.95\linewidth]{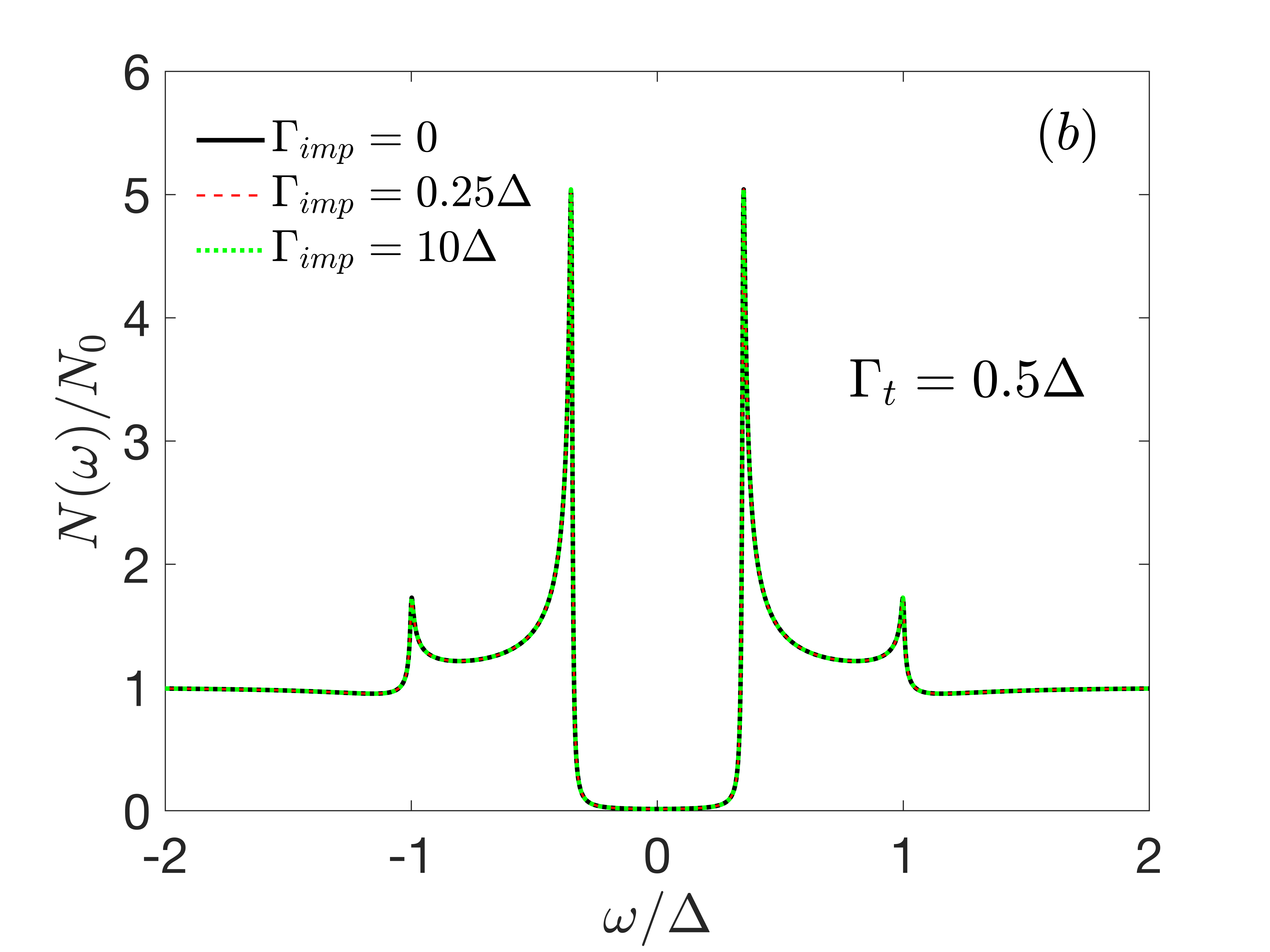}
\caption{Proximity induced gap in the normal metal. (a) Variation of normalized DOS with strength of tunneling energy scale $\Gamma_t$. (b) Normalized density of states at the interface as a function of energy for several values of impurity scattering rate.}
\label{Fig:Singlet}
\end{figure}

At this point, we can include the effect of impurity scattering. The impurity self-energy depends on the momentum integrated Green's function. Since $\mathbf{w}$ is an odd function of momentum, the terms linear in $\mathbf{w}$ vanish in the momentum integrated Green's function for a singlet superconductor. As a result, the  momentum integrated Green's function is,
\begin{eqnarray}
 \int_\mathbf{\mathbf{k}} \check{\mathbb{G}}  &=& -\pi N_0 \frac{1}{\sqrt{\Sigma_1^2-\bar{\omega}^2}} \begin{pmatrix}
 \bar{\omega} \sigma_0 & \Sigma_1 i \sigma_y  \\
 -i \sigma_y \Sigma_1 &  \bar{\omega} \sigma_0
 \end{pmatrix},
\end{eqnarray}
where $N_0$ is the normal metal DOS at the Fermi level.  The impurity self-energy can be expressed as,
\begin{eqnarray}
 \check{\Sigma}_{imp} &=&
\begin{pmatrix}
 {\Sigma}_{imp0}  \sigma_0 & i \sigma_y \Sigma_{imp1} \\
 -i \sigma_y \Sigma_{imp1} &  {\Sigma}_{imp0} \sigma_0,
 \end{pmatrix}, \\
 \Sigma_{imp0} &=& n_{imp}\frac{ g_0 V^2}{1- V^2 (g_0^2 - g_1^2)},  \label{Eq:Sigma0} \\
 \Sigma_{imp1} &=& n_{imp}\frac{-g_1 V^2}{1- V^2 (g_0^2 - g_1^2)},\label{Eq:Sigma1}
\end{eqnarray}
where we have neglected the $\Sigma_3$ component of the impurity self energy, which vanishes for a particle hole symmetric system. 
We can define an impurity renormalized energy and an induced off-diagonal self-energy as,
\begin{eqnarray}
\tilde{\omega}&=& \omega- \Sigma_0(\omega)  - \Sigma_{imp0}(\tilde{\omega}), \\
\tilde{\Sigma}_1 &=& \Sigma_1(\omega) + \Sigma_{imp1}(\tilde{\omega}).
\end{eqnarray}
Here, the fully dressed Green's function is used to calculate $\tilde{\omega}$, $\tilde{\Sigma}_1$ self-consistently. 
The equations for the renormalized $\tilde{\Sigma}_1$ and $\tilde{\omega}$ are identical as those for an $s$-wave superconductors with nonmagnetic impurities.
Now, we rewrite the impurity and tunneling dressed Green's function $\check{\mathbb{G}}$ by replacing $\bar{\omega}$ and $\Sigma_1$ with $\tilde{\omega}$ and $\tilde{\Sigma}_1$, respectively. 
 After some straightforward algebra, we get $\tilde{\omega}/\tilde{\Sigma}_1 = \bar{\omega}/\Sigma_1$. This ensures that the induced pairs remain robust against nonmagnetic disorder. Next, we consider the interface DOS given by
 \begin{eqnarray}
 N(\omega)=N_0 \mathrm{Im} \frac{\tilde{\omega}}{\sqrt{\tilde{\Sigma}_1^2-\tilde{\omega}^2}}.
 \end{eqnarray}
Fig. \ref{Fig:Singlet}(a) shows the interface DOS for several values of the tunneling energy scale $\Gamma_t$. In the weak tunneling regime ($\Gamma_t\ll \Delta$), the effective gap in the DOS is determined by $\Gamma_t^2/\Delta$, and in the strong tunneling regime ($\Gamma_t\gg \Delta$), the effective gap becomes identical to the size of the gap of the superconductor. The sub-dominant triplet component does not induce any low energy sub-gap states, instead the low energy  DOS is mainly controlled by the singlet order parameter, which is an isotropic $s$-wave in the present case. The effect of impurity scattering is depicted in Fig. \ref{Fig:Singlet}(b), which shows no change in the DOS with increasing impurity scattering rate $\Gamma_{imp}\equiv n_{imp}\pi N_0 V^2/(1+\pi^2 V^2 N_0^2) $. {The underlying triplet order parameter remains unaffected by the impurity scattering, in contrast to a bulk triplet superconductors.}

\subsection{SN junction with a triplet superconductor}\label{sec:triplet}
In a triplet superconductor the order parameter is $\hat{\Delta}=\Delta \mathbf{d}\cdot \boldsymbol{\sigma}i\sigma_y$, where the $\mathbf{d}$-vector describes the pair structure in  spin space. Here, we restrict ourselves to unitary pairing ($\mathbf{d}\times\mathbf{d}^\ast=0$) in the superconducting component of the heterojunction.  In this case, we find for the tunneling self-energy,
\begin{eqnarray}
\check{\Sigma}_t(k_{||},\omega) = |\gamma|^2 \begin{pmatrix}
\Sigma_0 \sigma_0 & {\Sigma}_1 \tilde{\mathbf{d}} \cdot \boldsymbol{\sigma}  i \sigma_y  \\
-i \sigma_y {\Sigma}_1 \tilde{\mathbf{d}} ^\ast \cdot \boldsymbol{\sigma}& \Sigma_0 \sigma_0 
\end{pmatrix},
\end{eqnarray}
where the components  are given by,
\begin{eqnarray}
\Sigma_0 & = &  |\gamma|^2 \int \frac{d k_\perp}{2 \pi}  \frac{\omega }{\omega^2-\xi^2-\Delta^2 | \mathbf{d}|^2}, \\
\Sigma_1 \tilde{\mathbf{d}} & = &  |\gamma|^2 \int \frac{d k_\perp}{2 \pi}  \frac{-\Delta \mathbf{d}}{\omega^2-\xi^2-\Delta^2 | \mathbf{d}|^2 }  .
\end{eqnarray}
As before in the singlet case, we  ignore the $\Sigma_3$ self-energy. The $\mathbf{d}$-vector is an odd function of momentum and $\mathbf{\Sigma}_1$ is also an odd function of the momentum parallel to the interface ($k_{||}$), and it does not depend on the momentum component normal to the interface ($k_\perp$). The Green's function in the normal metal can be written as,
\begin{eqnarray}
\check{\mathbb{G}} &=& \begin{pmatrix}
\left[ \bar{\omega} - {\xi} \right] \sigma_0 - \epsilon_N (\mathbf{w}  \cdot \boldsymbol{\sigma})   & - {\Sigma}_1 \tilde{\mathbf{d}} \cdot \boldsymbol{\sigma}  i \sigma_y \\
 i \sigma_y {\Sigma}_1 \tilde{\mathbf{d}} ^\ast \cdot \boldsymbol{\sigma}    & \left[\bar{\omega} + {\xi} \right] \sigma_0 - \epsilon_N (\mathbf{w}  \cdot \boldsymbol{\sigma}^\ast) 
\end{pmatrix}^{-1} \nonumber \\
&=& \begin{pmatrix}
\hat{\mathbb{G}}_{11} & \hat{\mathbb{G}}_{12} \\
\hat{\mathbb{G}}_{21} & \hat{\mathbb{G}}_{22}.
\end{pmatrix}
\end{eqnarray}
The general structure of the off-diagonal Green's function has the form,
\begin{eqnarray}
 \hat{\mathbb{G}}_{12}  &\propto& A_0 \sigma_0 + A_1  \mathbf{w}\cdot\boldsymbol{\sigma} + A_2 \tilde{\mathbf{d}} \cdot\boldsymbol{\sigma}  + A_3 \tilde{\mathbf{d}} ^\ast\cdot\boldsymbol{\sigma}  \nonumber \\
 & & + A_4 \left( \tilde{\mathbf{d}} \times \mathbf{w} \right) \cdot\boldsymbol{\sigma},
 \label{Eq:Trp_G12g}
\end{eqnarray}
where the specific values of the set $\{A_i\}$ (i=0,...4) will depend on geometry of the junction and the specifics of the $\mathbf{d}$-vector which in turn determine $\tilde{\mathbf{d}}$.
In general,  $\tilde{\mathbf{d}}  \times \tilde{\mathbf{d}} ^\ast$ may not vanish despite unitary pairing in the superconductor ($\mathbf{d}\times\mathbf{d}^\ast=0$). 
Eq. \eqref{Eq:Trp_G12g} has singlet and triplet terms, but due to the anisotropic nature of the gap in the superconductor details of geometry and pairing structure is essential to understand the structure of induced pairs in the normal metal. To proceed, we will consider a few pertinent cases.

\subsubsection{$\mathbf{d}$-vector $|| \mathbf{w}$}\label{sec:R2}
\begin{figure}
\includegraphics[width=.95\linewidth]{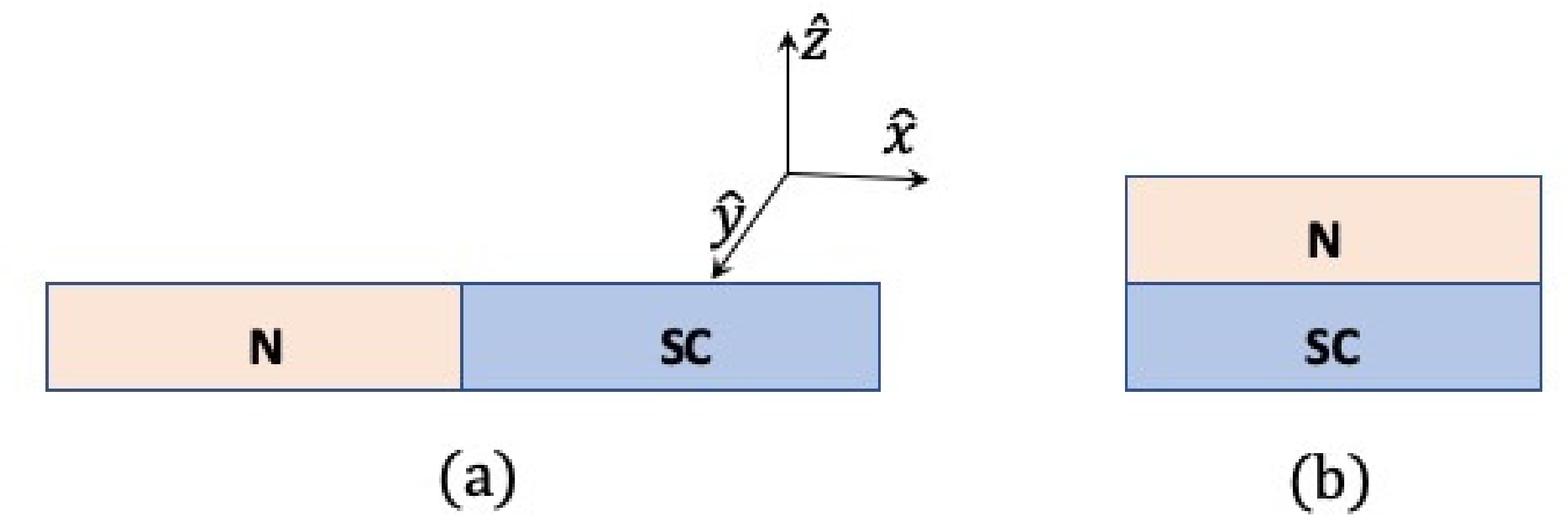}
\caption{Two different SN junction geometries.}
\label{Fig:SNG}
\end{figure}

Motivated by the recent experiments on CoSi$_2$/TiSi$_2$/Si heterostructures\cite{Chiu2021,Mishra2021,Chiu2022}, we first consider the case of a $\mathbf{d}$-vector that is parallel to the SOC vector $\mathbf{w}$.  The experimental results for CoSi$_2$/TiSi$_2$ on a Si substrate indicate the presence of a dominant triplet superconducting state in CoSi$_2$ with $\mathbf{d}$-vector along $\mathbf{w}$.  Due to the SOC which is induced by the Si substrate, there will also be a finite but weak  singlet component in the superconductor which we ignore for now. The general case of a mixed parity superconductor will be addressed in subsection \ref{Sec:mx}. 
For side-by-side coupled junctions as the one illustrated in Fig.~\ref{Fig:SNG}(a), the tunneling self-energies are,
\begin{eqnarray}
\Sigma_0 &=& \Sigma_0(\omega,k_{||}), \\
\Sigma_1 \tilde{\mathbf{d}} &=& - k_y \Sigma_{1}(\omega,k_{||}) \hat{\mathbf{x}}  ,
\label{Eq:TSE_tr_sbs}
\end{eqnarray}  
where $\tilde{\mathbf{d}}  = -k_y \hat{\mathbf{x}}$ and the normal and anomalous part of the Green's functions are,
\begin{eqnarray}
\hat{G}_{11} &=& \frac{1}{D} \left[ \left( a_0 b_+ b_- - b_0\Sigma_1^2 \tilde{\mathbf{d}}\cdot \tilde{\mathbf{d}} \right) \sigma_0 \right. \nonumber \\
& &- \epsilon_N \left( b_+ b_- +\Sigma_1^2 \tilde{\mathbf{d}}\cdot \tilde{\mathbf{d}}  \right)(\mathbf{w}  \cdot \boldsymbol{\sigma})  \nonumber \\
 && \left. -2\epsilon_N \Sigma_1^2 (\tilde{\mathbf{d}}\cdot \mathbf{w}) \tilde{\mathbf{d}}  \cdot \boldsymbol{\sigma}  \right] \label{Eq:G11_trp_sbs} \\
\hat{G}_{12} &=& \hat{\mathtt{G}}_{12}\frac{1}{D} i \sigma_y \\
\hat{\mathtt{G}} _{12}&=&  (2 \epsilon_N \xi_\mathbf{k} {\Sigma}_1 \tilde{\mathbf{d}}  \cdot \mathbf{w} ) \sigma_0 \nonumber \\
&& -2 \epsilon_N^2 {\Sigma}_1 (\tilde{\mathbf{d}} \cdot \mathbf{w}) \mathbf{w}  \cdot \boldsymbol{\sigma}  \nonumber \\
&&+    ( \bar{\omega}^2 -\xi_{\mathbf{k}}^2 +\epsilon_N^2  -{\Sigma}_1^2 \tilde{\mathbf{d}}\cdot \tilde{\mathbf{d}}  ) {\Sigma}_1 \tilde{\mathbf{d}}  \cdot \boldsymbol{\sigma}  \nonumber \\
&&-2 i \bar{\omega} \epsilon_N   {\Sigma}_1 (\tilde{\mathbf{d}} \times \mathbf{w} ) \cdot \boldsymbol{\sigma}  \label{Eq:G12_trp_sbs} \\
D&=& (\bar{\omega}^2 -\xi_{\mathbf{k}}^2+\epsilon_N^2-{\Sigma}_1^2 \tilde{\mathbf{d}}\cdot \tilde{\mathbf{d}}  )^2 \nonumber \\
&& + 4 \epsilon_N^2\left[  {\Sigma}_1^2 (\tilde{\mathbf{d}} \cdot \mathbf{w})^2 -\bar{\omega}^2\right].
\end{eqnarray}

Here $a_0 = \bar{\omega}-\xi_\mathbf{k}$,  $b_0 = \bar{\omega}+\xi_\mathbf{k}$ and  $b_\pm$ is  $\bar{\omega}+\xi_\mathbf{k}\pm \epsilon_N$. Eq. \eqref{Eq:G12_trp_sbs} contains an even parity ($\propto k_y^2$), even frequency singlet term that arises due to a non-zero SOC and disappears in the limit of vanishing SOC. Apart 
from an expected triplet pairing with $\tilde{\mathbf{d}} \cdot\boldsymbol{\sigma}$, a non-vanishing SOC brings about another kind of triplet pairing with $\mathbf{w}\cdot\boldsymbol{\sigma}$ 
structure.  Both these triplet components are even in frequency and have odd parity. The novel pairing that arises due to the SOC is described by the last term in Eq.  \eqref{Eq:G12_trp_sbs} which is $\propto \left( \tilde{\mathbf{d}}  \times \mathbf{w} \right)\cdot\boldsymbol{\sigma}$ and possesses a momentum dependence $\propto k_x k_y$. 
This term describes triplet pairs that are have even parity and are odd in  frequency. This term is absent when the SOC vanishes.  This term also  vanishes for junctions possessing  top-bottom geometry as the one shown in Fig. \ref{Fig:SNG}(b) as  $\tilde{\mathbf{d}}  || \mathbf{w}$ in this geometry. Thus, there are only singlet and triplet components with even frequency structure in this junction geometry where the singlet pairs are generated by the SOC.
\begin{figure}
\includegraphics[width=.95\columnwidth]{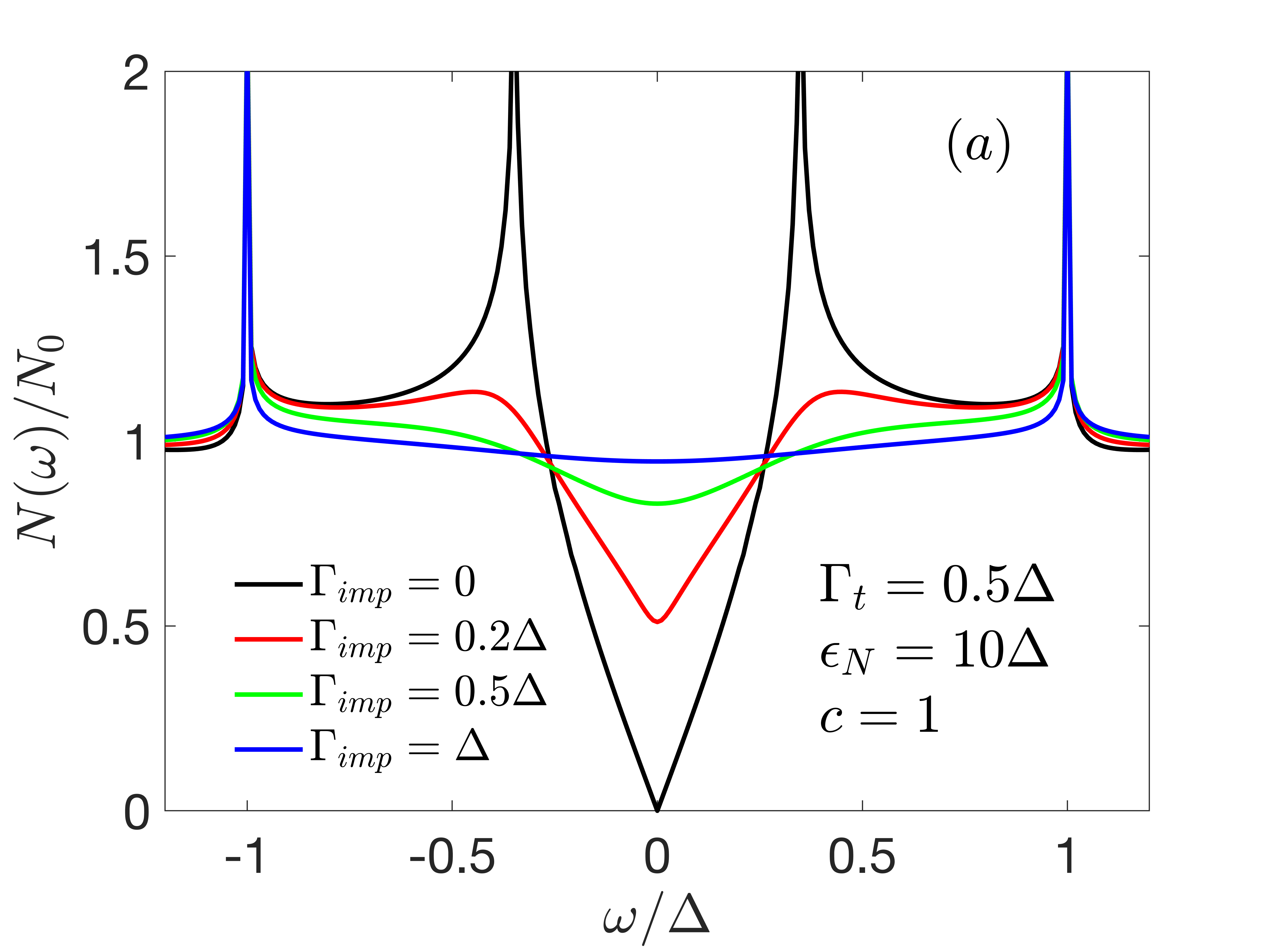}
\includegraphics[width=.95\columnwidth]{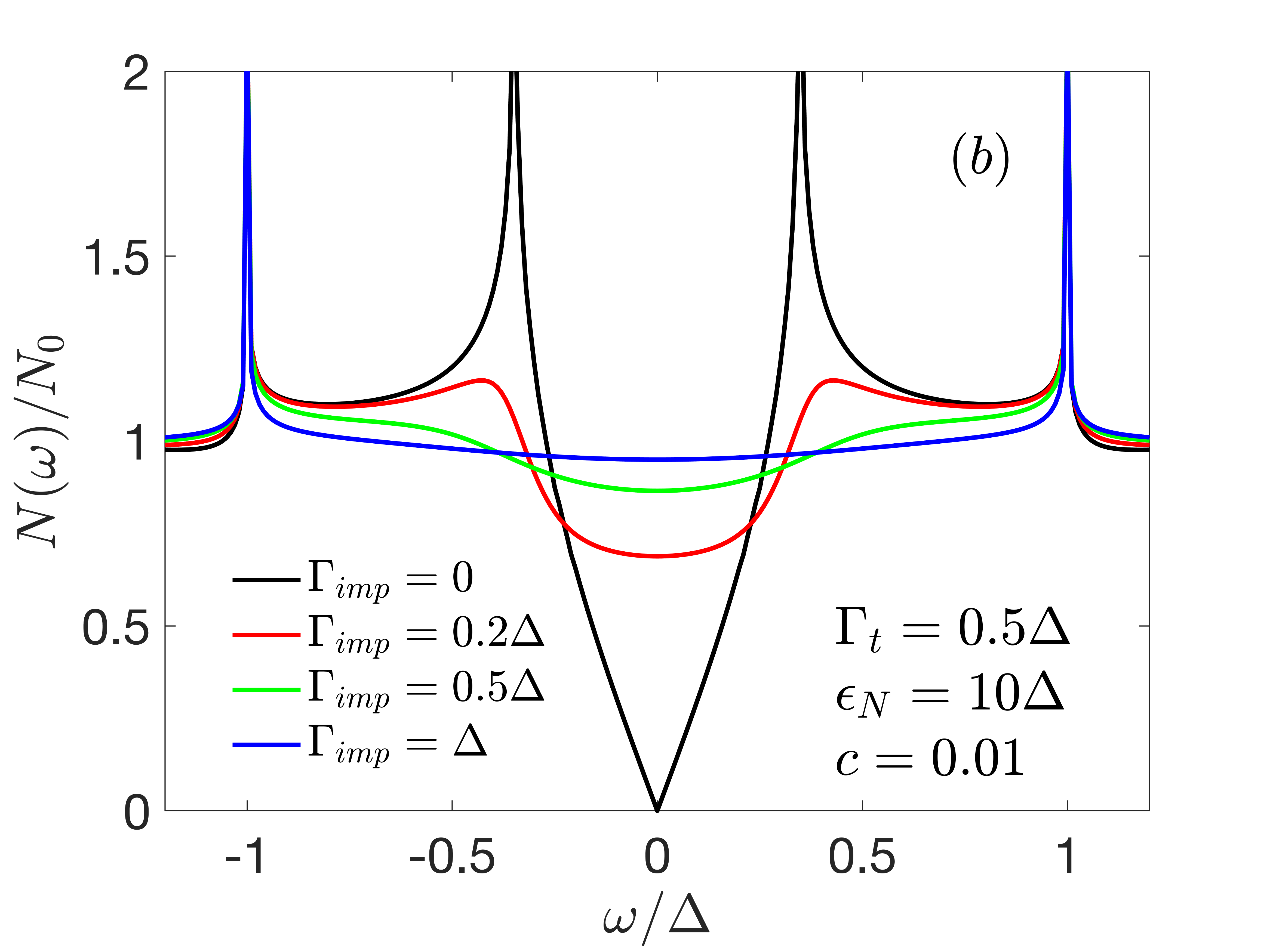}
\caption{The local density of states at the interface for several values of normal state scattering rate. The parameter $c$ is the cotangent of $s$-wave  phase shift $c\equiv \cot \theta_s$ is $1$ for panel (a) and $0.01$ for panel(b). }
\label{Fig:triplet1_sbs_dos}
\end{figure}
Next, we include the effect of impurity scattering as described in Sec. \ref{sec:MF}. We consider a two-dimensional electron gas and a side-by-side coupled geometry. For a two dimensional superconductor with an order parameter characterized by $\mathbf{d}=\hat{\mathbf{z}}\times \mathbf{k}$, the Fermi surface is fully gapped. In that case, the self-energy components are
\begin{eqnarray}
\Sigma_0 = -\Gamma_t \frac{\omega}{\sqrt{\Delta^2-\omega^2}}, \\
{\Sigma}_1 \tilde{\mathbf{d}}= \Gamma_t \frac{\Delta}{\sqrt{\Delta^2-\omega^2}} \frac{k_y}{k_F} \hat{\mathbf{x}}.
\end{eqnarray}
The momentum integrated Green's function is $g_0 \check{{1}}$. The disorder renormalized $\tilde{\omega}$ is determined by,
\begin{eqnarray}
\tilde{\omega}&=& \omega + \Gamma_t \frac{\omega}{\sqrt{\Delta^2 - \omega^2}}  +\frac{n_{imp}}{\pi N_0} \frac{\mathtt{g}_0}{\cot^2 \theta_s- \mathtt{g}_0^2}
\end{eqnarray}
where $\theta_s\equiv \tan^{-1} (\pi N_0 V) $ is the $s$-wave scattering phase shift. Fig. \ref{Fig:triplet1_sbs_dos}(a) shows the local DOS at the interface for weak scattering ($c=1$)  and panel (b) shows the local DOS for $c=0.01$, {\itshape i.e.} strong scattering. In contrast to isotropic $s$-wave,  the impurity scattering  rapidly suppresses the induced superconductivity in this case. While the {bulk} superconductor is fully gapped, the induced superconductivity has low-energy states and does not develop a gap in the DOS. {Although, the low energy states rapidly disappear with increasing disorder, the off-diagonal Green's function remains finite. This is in contrast to a bulk superconductor, where  the anomalous Green's function vanishes once the impurity scattering rate reaches a critical value.}
\begin{figure}
\includegraphics[width=.95\linewidth]{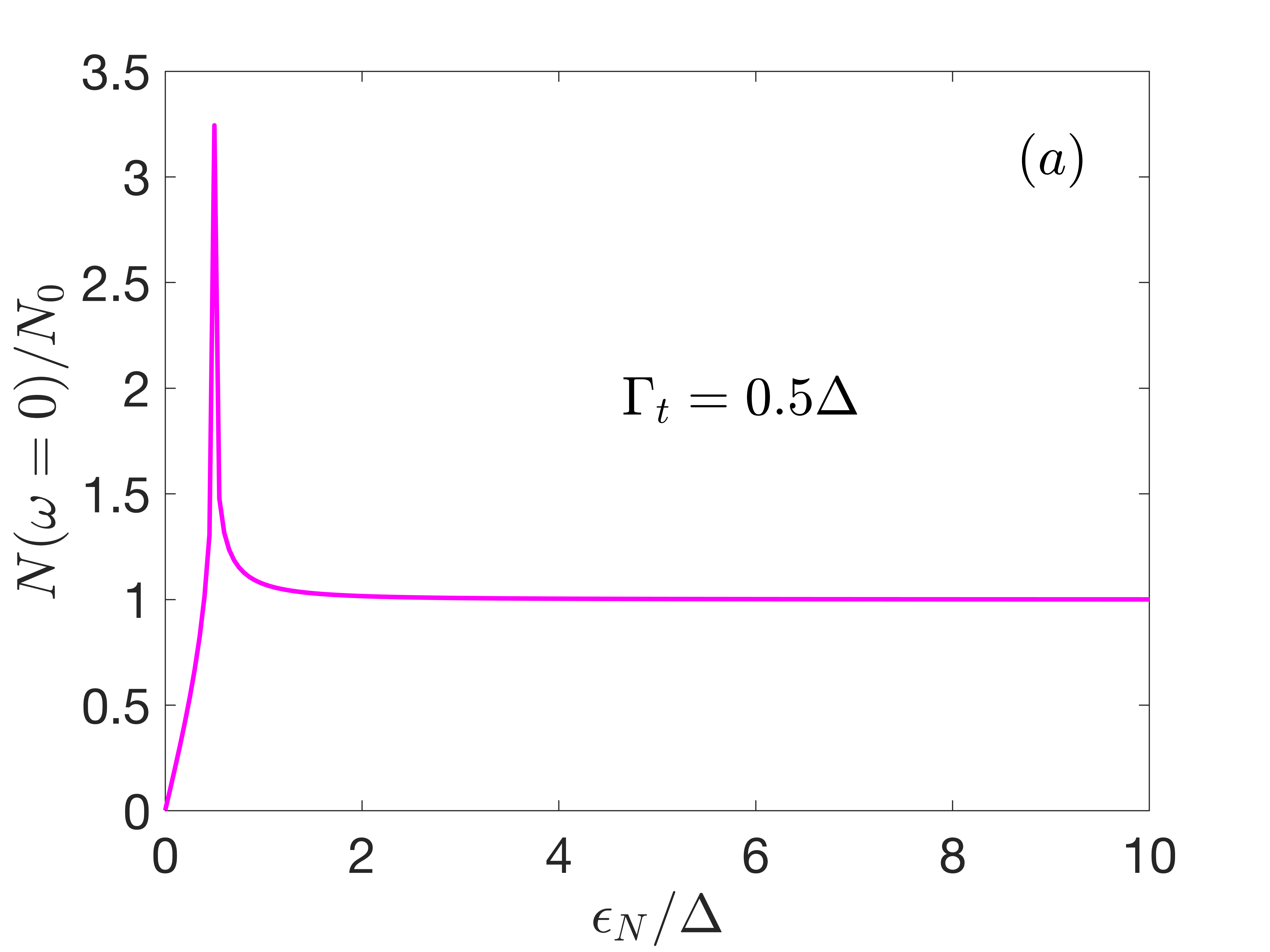}
\includegraphics[width=.95\linewidth]{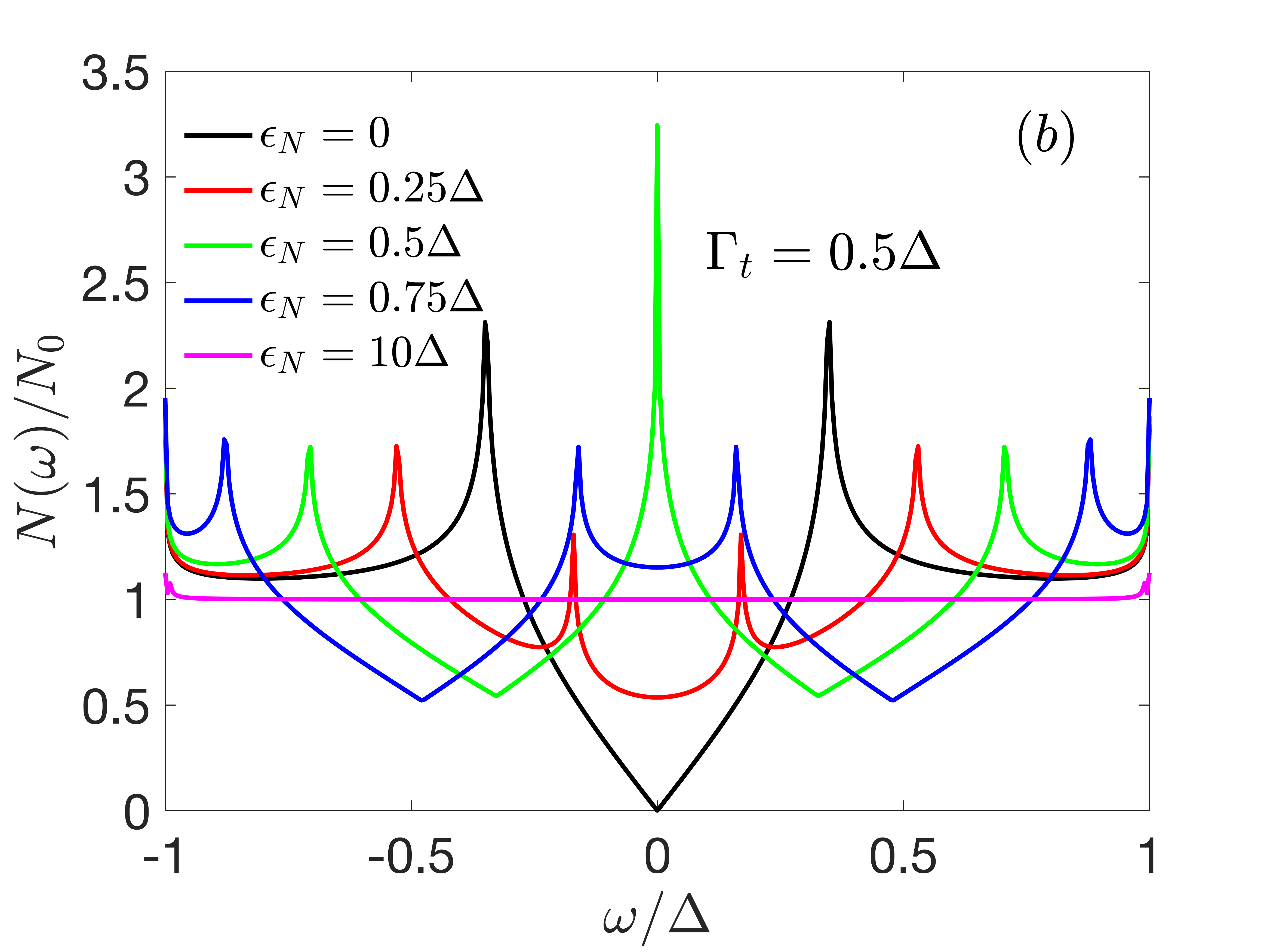}
\includegraphics[width=.95\linewidth]{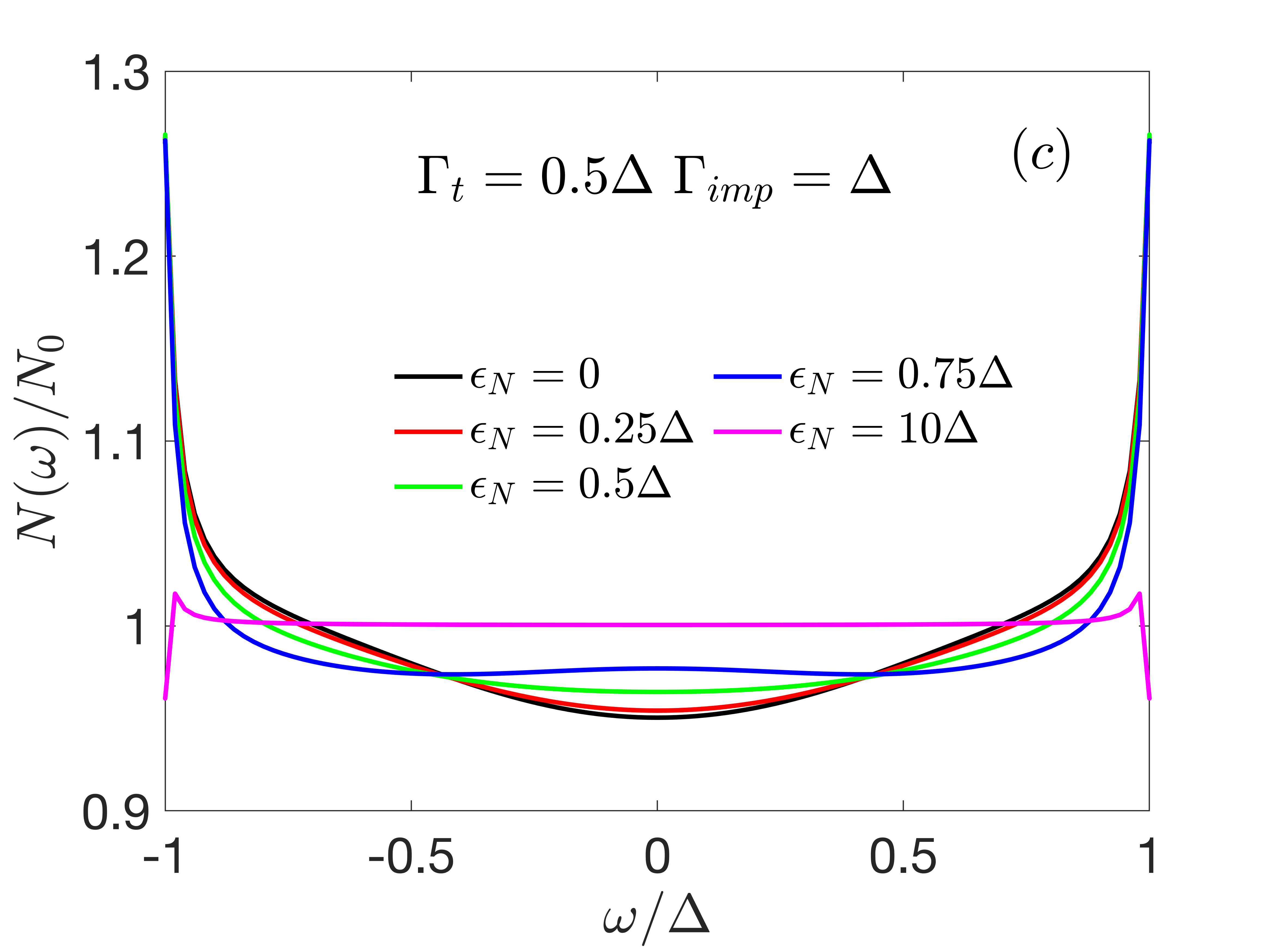}
\caption{(a) The interface DOS at the Fermi level as a function of SOC energy in the clean limit. The interface DOS as a function of energy for several values of SOC energy in the clean limit (b) and with a scattering rate $\Gamma_N=\Delta$ for $c=1$.}
\label{Fig:chiral_sbs_dos}
\end{figure}

\subsubsection{$\mathbf{d}$-vector $\perp \mathbf{w}$}
Next, we consider a chiral $p$-wave state with $\mathbf{d} =( p_x + i p_y)\hat{\mathbf{z}}$ which is a complex, but unitary order parameter ($\mathbf{d}\times \mathbf{d}^\ast=0$). In this case, the tunneling self-energies read,
\begin{eqnarray}
\Sigma_0 &=& \Sigma_0(\omega,k_{||}), \\
 \Sigma_1 \tilde{\mathbf{d}}  &=& i k_y \Sigma_{1}(\omega,k_{||}) \hat{\mathbf{z}},
\label{Eq:TSE_tr_chiral_p}
\end{eqnarray}  
with $\tilde{\mathbf{d}}=i k_y \hat{\mathbf{z}}$. The anomalous part of the Green's function is obtained as,
\begin{eqnarray}
\hat{G}_{11} &=&\left[ \left( a_0 b_+ b_- - b_0\Sigma_1^2 | \tilde{\mathbf{d}}|^2\right) \sigma_0 \right. \nonumber \\
& & \left. + \epsilon_N \left( b_+ b_- +\Sigma_1^2 | \tilde{\mathbf{d}} |^2 \right)(\mathbf{w}  \cdot \boldsymbol{\sigma}) \right]\frac{1}{D}  , \\ \label{Eq:G11_chip_sbs}
\hat{G}_{12} &=& \frac{ \hat{\mathtt{G}}_{12}  }{D} i \sigma_y. \label{Eq:G12_chip_sbs}  \\
\hat{\mathtt{G}}_{12} &=&  \Sigma_1  ( \bar{\omega}^2 -\xi_{\mathbf{k}}^2 +\epsilon_N^2  -{{\Sigma}}_1^2  |\tilde{\mathbf{d}} |^2  )  {\tilde{\mathbf{d}}} \cdot {\boldsymbol{\sigma}}  \nonumber  \\ 
& & -2i \bar{\omega} \epsilon_N   {\Sigma}_1( {\tilde{\mathbf{d}}}\times \mathbf{w} ) \cdot \boldsymbol{\sigma} 
\\
D&=& (\bar{\omega}^2 -\xi_{\mathbf{k}}^2+\epsilon_N^2-{{\Sigma}_1}^2 | \tilde{\mathbf{d}} |^2 )^2
- 4 \epsilon_N^2 \bar{\omega}^2.
\end{eqnarray}

As $\tilde{\mathbf{d}} \perp \mathbf{w}$, any $\tilde{\mathbf{d}}\cdot\mathbf{w}$ component be it singlet or triplet, vanishes. This also ensures the presence of an odd frequency component for both kinds of junction geometries. 
The  momentum structure of the odd frequency component turns out to be $\propto k_x k_y \hat{\mathbf{x}} + k_y^2 \hat{\mathbf{y}}$ therefore, the $\Sigma_{imp1}$ self-energy contribution becomes finite, in contrast to the previous case.  A non-vanishing $\Sigma_{imp1}$ self-energy, as it turns out, converts the induced pairing into non-unitary pairing. (See the general expression for the Green's function in the appendix \ref{App1}).
The impurity self-energies are given by Eq. \eqref{Eq:Sigma0} and \eqref{Eq:Sigma1}. However, the SOC itself kills the induced pairing  as can be inferred from  Fig. \ref{Fig:chiral_sbs_dos}(a), where it is shown that the DOS at the Fermi level reaches the normal state value as the SOC strength increases beyond the pairing energy scale of the superconductor. In Fig. \ref{Fig:chiral_sbs_dos}(b), the DOS at the interface is shown for energies below the superconducting gap in the case of weak SOC. 
We are, however, interested in the regime $\Delta \ll \epsilon_N \ll E_F$, and in this regime the proximity induced superconductivity does not survive. {In the strong SOC regime, unless the $\mathbf{d}$-vector is aligned with SOC, SOC will act as a strong pair-breaker\cite{Frigeri2004}}. 
The inclusion of impurity scattering does not change this conclusion but further diminishes the proximity induced superconductivity even for weak SOC as  be inferred from Fig. \ref{Fig:chiral_sbs_dos}(c).
\begin{figure}[b]
\includegraphics[width=.99\columnwidth]{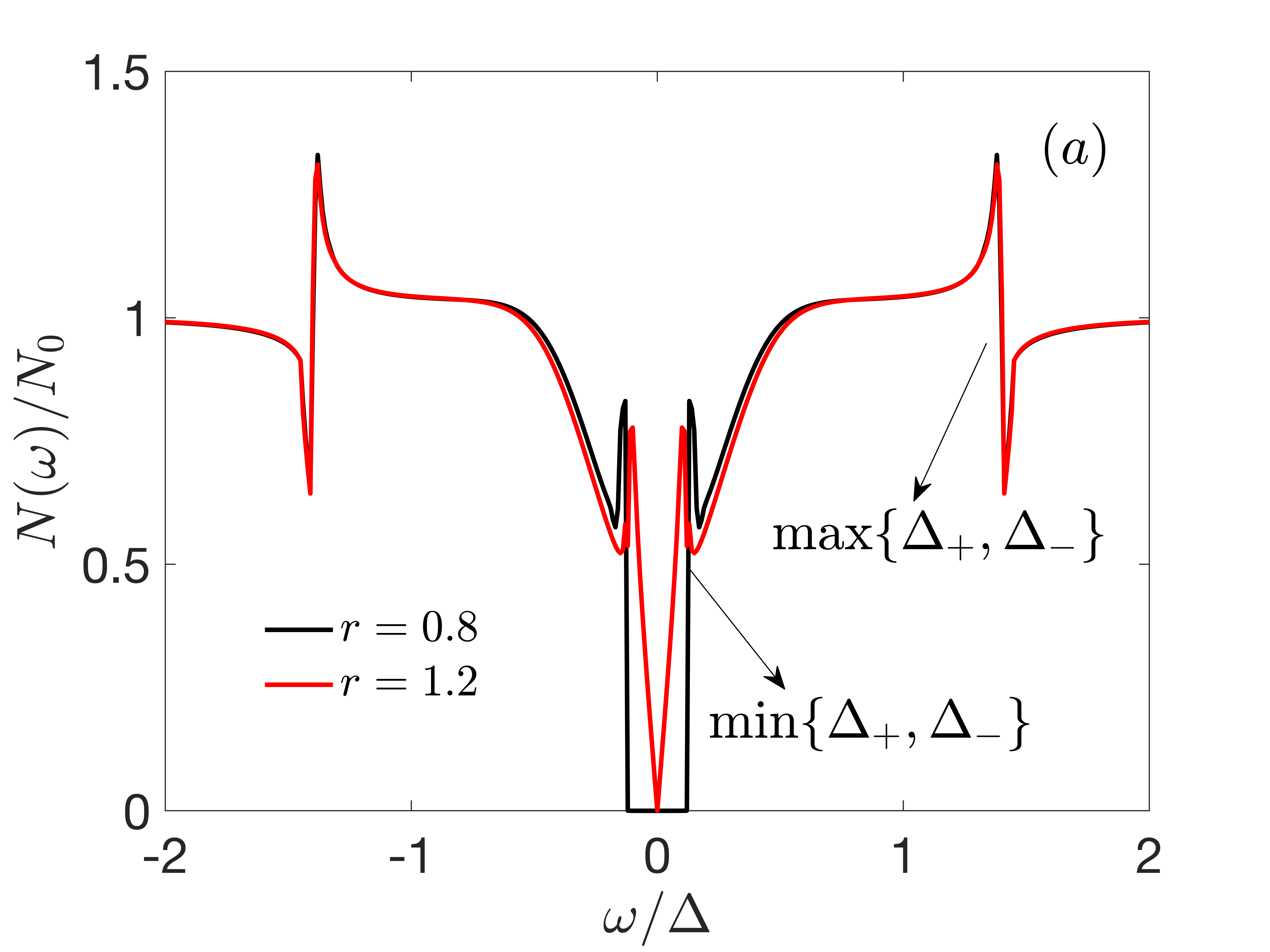}
\includegraphics[width=.99\columnwidth]{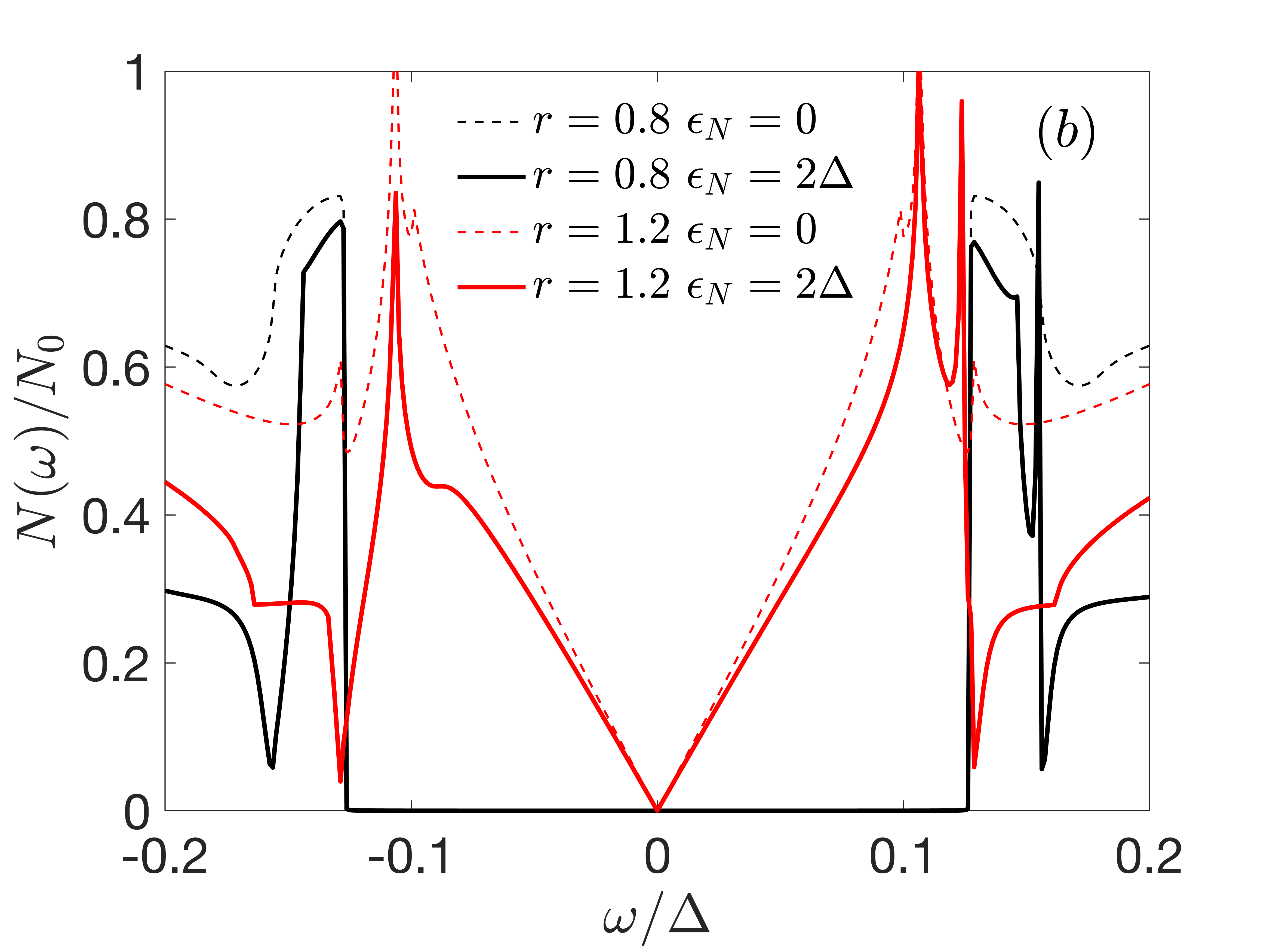}
\caption{The interface DOS for a mixed parity and normal junction without any impurity scattering. (a) The interface DOS for $r=0.8$ and $r=1.2$ with no SOC in the normal metal. The tunneling energy scale $\Gamma_t=0.5\Delta_0$. (b) The interface DOS states for $r=0.8$ and $r=1.2$ with (solid lines) and without (dashed lines). }
\label{Fig:mixed_soc_pha}
\end{figure}
\subsubsection{Mixed parity state}\label{Sec:mx}
The last case that we consider is that of a superconductor with mixed parity order  parameters. Such a state is possible, if the superconductor itself is under the influence of a SOC. The normal and anomalous  mean field Green's functions in this case are,
\begin{eqnarray}
\hat{G}_{11} &=&  \frac{1}{2} \left[ \hat{G}_+ + \hat{G}_-  \right] + \frac{1}{2} \left[ \hat{G}_+ - \hat{G}_-  \right]  \mathbf{d}\cdot \boldsymbol{\sigma}, \\
\hat{G}_{\pm} &=& \frac{\omega +\xi_\pm }{\omega^2 - \xi_\pm^2 -\Delta^2_\pm}, \\
\hat{G}_{12} &=& \frac{1}{2}\left[ (\hat{F}_+ + \hat{F}_-) + (\hat{F}_+ - \hat{F}_-) \mathbf{d}\cdot \boldsymbol{\sigma} \right] i \sigma_y, \\
\hat{F}_\pm &=& \frac{\Delta_\pm}{\omega^2 - \xi_\pm^2 -\Delta^2_\pm}.
\end{eqnarray}
Here $\Delta_\pm = (\Delta_s \pm \Delta_t)$ and $\xi_\pm=\xi_{\mathbf{k}}\pm \epsilon_S$, where $\epsilon_S$ is the SOC energy scale in the superconductor and $\Delta_{s/t} $ is the singlet/triplet component of the gap. The normal ($\hat{\Sigma}_{11}$) and  anomalous ($\hat{\Sigma}_{12}$) tunneling self-energies are $\Sigma_0 + \Sigma_{soc} \tilde{\mathbf{d}}\cdot \boldsymbol{\sigma}$ and $ \left( \Sigma_s + \Sigma_t \tilde{\mathbf{d}}\cdot \boldsymbol{\sigma} \right) i\sigma_y$, where $\Sigma_{soc}$ modifies the SOC in the normal metal and $\Sigma_{s/t}$  is the singlet/triplet component. They are given by 
\begin{eqnarray}
\Sigma_{0/soc} &=& -\Gamma_t \left( \frac{\omega}{Q_+} \pm \frac{\omega}{Q_-}\right), \\
\Sigma_{s/t} &=& \Gamma_t   \left( \frac{\Delta_+}{Q_+} \pm \frac{\Delta_-}{Q_-}\right). 
\end{eqnarray}
where $Q_\pm = 2\sqrt{\Delta_\pm^2-\omega^2}$ and it was assumed that the SOC splitting energy in the superconductor is $\epsilon_S \ll E_F$.  Consequently, we ignore the differences between the DOS of the two helical bands, which is of the order of $\epsilon_S/E_F$.  The renormalized SOC term in the normal metal can be rewritten as,
\begin{eqnarray}
\tilde{\epsilon}_N \widetilde{\mathbf{w}}\cdot \boldsymbol{\sigma} &=& \epsilon_N \mathbf{w}\cdot \boldsymbol{\sigma} +\Sigma_{soc} \tilde{\mathbf{d}}\cdot \boldsymbol{\sigma},\\
\tilde{\epsilon}_N &=&  \sqrt{(\epsilon_N \mathbf{w}+\Sigma_{soc} \tilde{\mathbf{d}})\cdot(\epsilon_N \mathbf{w}+\Sigma_{soc} \tilde{\mathbf{d}})},\\
\widetilde{\mathbf{w}}&=& \frac{\epsilon_N \mathbf{w} + \Sigma_{soc} \tilde{\mathbf{d}}}{\tilde{\epsilon}_N}.
\end{eqnarray}
The anomalous Green's function in the normal metal is,
\begin{eqnarray}
\hat{G}_{12} &\propto &\left[ A_0 \sigma_0 + A_1 \widetilde{\mathbf{w}}\cdot \boldsymbol{\sigma} + A_2 \tilde{\mathbf{d}}\cdot \boldsymbol{\sigma} \right. \nonumber \\
& &  \left. + A_3 ( \tilde{\mathbf{d}} \times \widetilde{\mathbf{w}}  )\cdot \boldsymbol{\sigma}   \right]i\sigma_y,
\label{Eq:G12_mixed}
\end{eqnarray}
where the  set  $\{A_i\}~(i=0,\dots,3)$ of coefficients is provided in appendix \ref{App2}.
The first term in Eq. \eqref{Eq:G12_mixed} is the even parity, spin singlet term and the second and third terms are odd parity, spin triplet terms. There is an odd frequency,  spin triplet, even parity (OTE) term, which only exists in the presence of both a finite triplet component in the superconductor and a finite SOC in the normal metal.
\begin{figure}[b]
\includegraphics[width=.95\columnwidth]{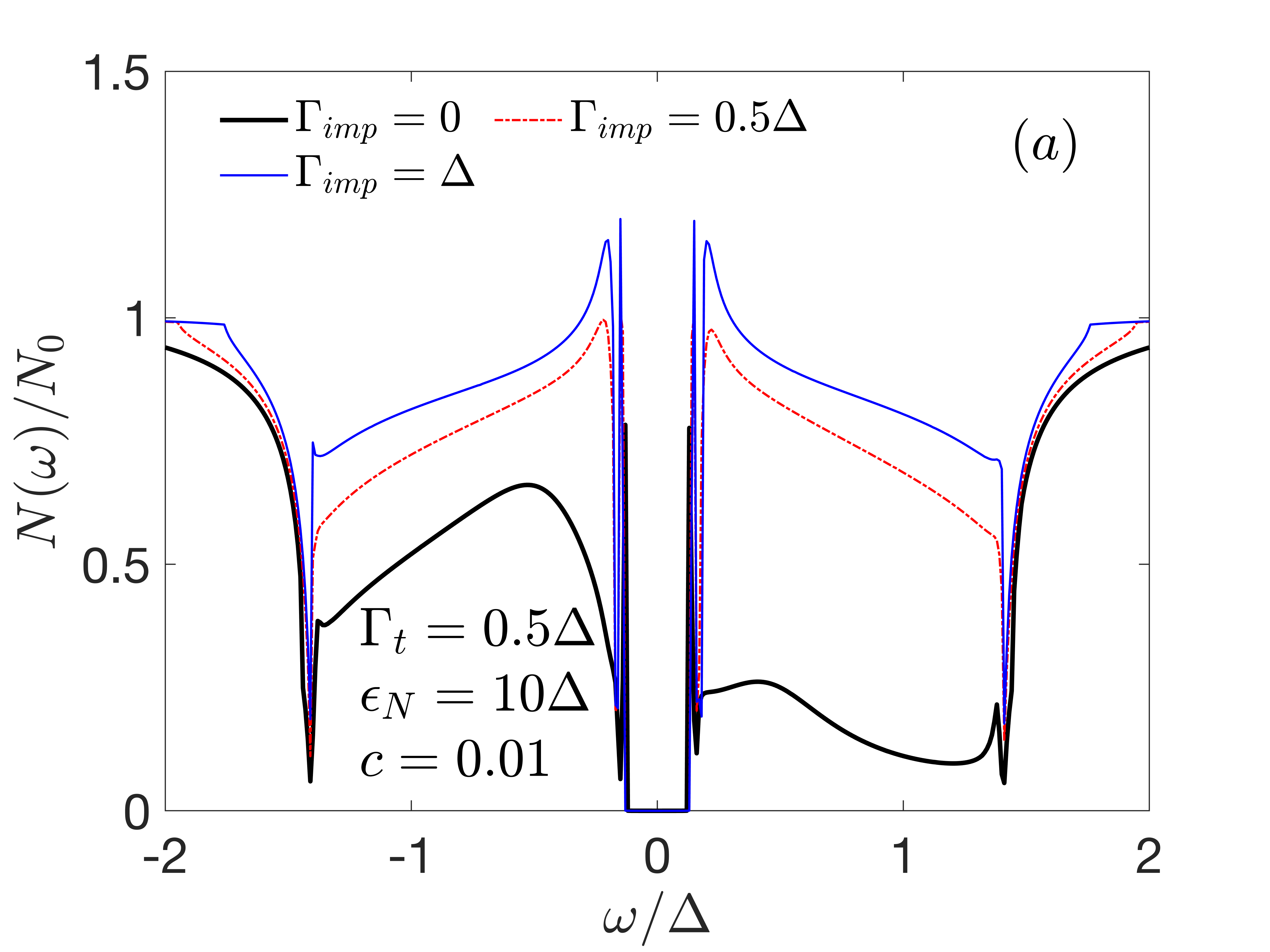}
\includegraphics[width=.95\columnwidth]{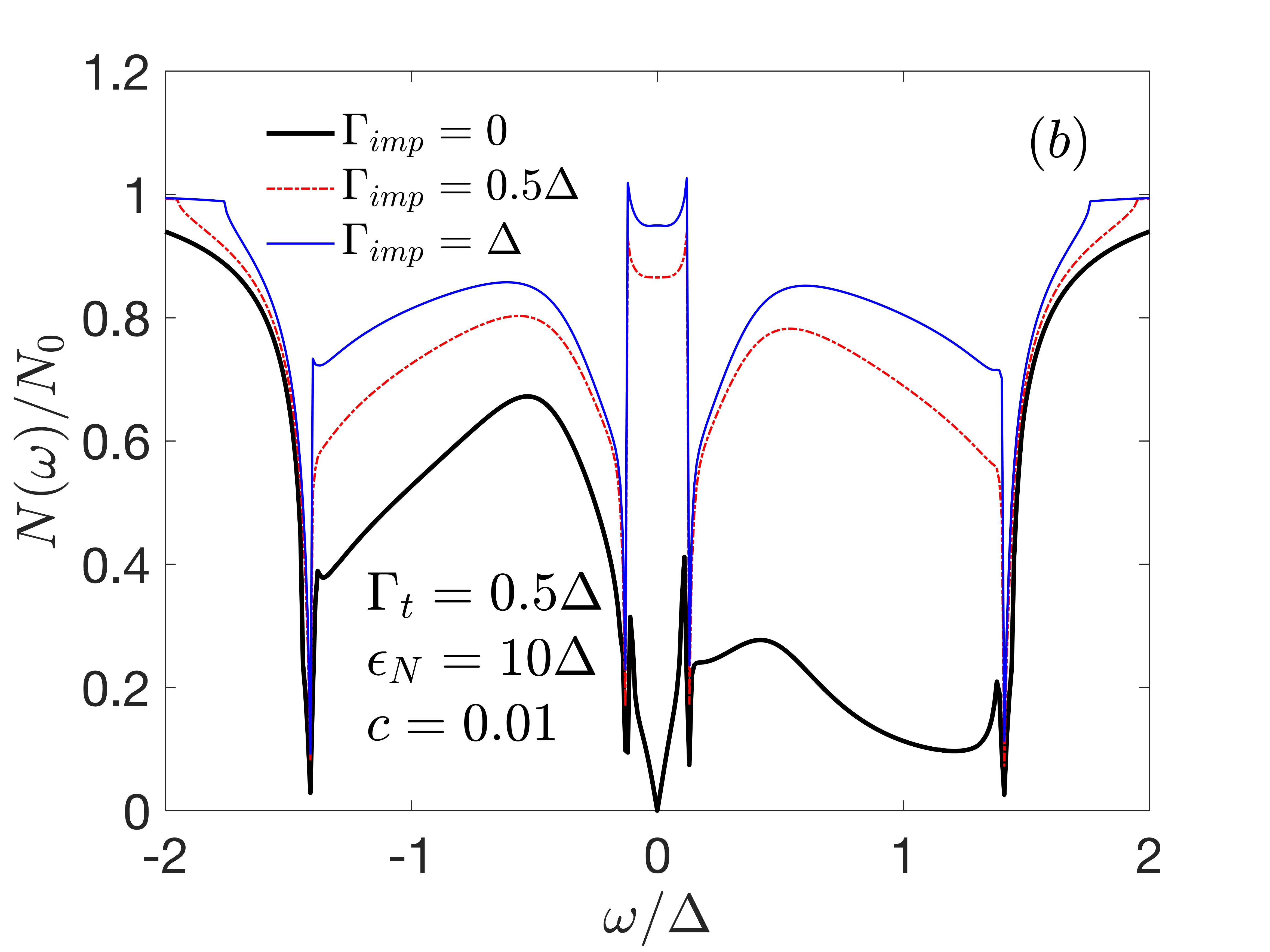}
\caption{(a) The interface DOS for singlet dominant ($r=0.8$) SN junction for several representative values of impurity scattering rate. (b) Variation of interface DOS with disorder for triplet dominant ($r=1.2$) case. }
\label{Fig:mixed_soc_dis}
\end{figure}

We parameterize the singlet and the triplet gaps in the superconductor as $\Delta_s = \Delta_0/\sqrt{1+r^2}$ and $\Delta_t = \Delta_0 r/\sqrt{1+r^2}$, respectively\cite{Mishra2021}.
The parameter $r$ is the ratio of triplet and singlet order parameters, where $r>1$ describes a triplet dominant  regime while $r<1$ presents a singlet dominant regime. For a mixed parity superconductor such as non-centrosymmetric superconductors any value of $r \in [0,\infty)$ is possible.   Its  experimental determination typically relies on  superfluid density or thermal transport measurements. In principle, SN and proximity junctions can also be used to determine the relative strength of singlet and triplet component \cite{Chiu2021,Mishra2021,Chiu2022}. In certain systems, $r$ appears to be tunable \cite{Ishihara2021}. The general concerns regarding the determination of the order parameter and its symmetry apply here as well. A consistent extraction of similar $r$ values from different experiments is appears as the most promising strategy.

As is clear from the defining expression, in the extreme singlet ($r\rightarrow 0$) and triplet ($r\rightarrow \infty$)  limit,  we recover the results presented in  Sec. \ref{sec:R1} and  \ref{sec:R2}, respectively (See also appendix \ref{App3}). In the intermediate regime, the DOS at the interface shows signatures of two energy scales $\Delta_\pm = \Delta_s \pm \Delta_t$. For a vanishing SOC in the normal metal, the singlet dominant regime ($r<1$) ensues, where the DOS shows a gap $~\mathrm{min}\lbrace \Delta_+, \Delta_-\rbrace$, as shown Fig. \ref{Fig:mixed_soc_pha}(a) and for the triplet dominant case ($r>1$), the DOS is similar to the pure triplet case discussed above (see Sec. \ref{sec:R2}), but the relevant energy scale is again $\mathrm{min}\lbrace \Delta_+, \Delta_-\rbrace$. This behavior remains qualitatively same in the presence of a SOC in the normal metal.  However, the SOC leads to a particle-hole asymmetric DOS as illustrated in Fig. \ref{Fig:mixed_soc_pha}(b). This happens due to the coupling of  singlet and triplet pairs in the normal metal due to the SOC.  This coupling generates a term linear in $\xi$ in the denominator of the Green's function, with a prefactor $\propto \tilde{\epsilon}_N \Sigma_s \Sigma_t  \tilde{\mathbf{d}}\cdot \widetilde{\mathbf{w}}$, which causes the breakdown of particle-hole symmetry.
The effect of  impurity scattering for the singlet/triplet dominant case is qualitatively the same as for the pure singlet/triplet case. 
The particle-hole  asymmetry in the DOS also gets smeared by the impurity scattering as depicted in  Fig. \ref{Fig:mixed_soc_dis}(a) for the singlet dominant ($r<1$) case and in the Fig. \ref{Fig:mixed_soc_dis}(b) for the triplet dominant case ($r>1$).

\section{Summary \& Conclusion}
In this paper, we have studied the effect of  Rashba SOC on the structure of proximity induced Cooper pairs in a normal metal  connected to a superconductor.
We considered several kinds of gap symmetries.
The SOC in the normal metal leads to a singlet-triplet mixed state if the SN junction involves a singlet superconductor. The strength of the triplet component in this case  depends on the strength of the SOC and the low energy quasiparticle spectrum remains gapped and robust against disorder. The SOC driven triplet state does not lead to any low energy states. This is reminiscent of the  proximity induced mixed parity states that have been reported for topological insulator and $s$-wave superconducting junctions where the origin  of the singlet-triplet mixing is  the spin-momentum locking \cite{FuKane2008,Stanescu2010,Lababidi2011,Potter2011,*Potter2011Err,Yokoyama2012,YuWu2016}. 

In the case of an SN junction involving a triplet superconductor, the broken spin-rotational symmetry can lead to a non-zero singlet component, but its presence  depends on the spin-structure of the gap in the superconductor and the junction geometry.  We find that a singlet component is present whenever the off-diagonal self-energy ($\mathbf{\Sigma}_1$) has a component parallel to the SOC vector ($\mathbf{w}$). On the other hand, an off-diagonal self-energy that is perpendicular to the SOC vector get suppressed by the SOC very quickly.
The induced triplet component may have a different spin-structure compared to the superconductor, depending on the junction geometry. The induced triplet pairs have sub-gap low energy states, however, the induced triplet component turns out to be fragile against impurity scattering. The effect of disorder is similar to the effect of disorder on proximity induced $p$-wave superconductivity on the surface of  topological insulators\cite{Tkachov2013}. In the SN junctions with two component superconductors that have both singlet and triplet components, the low energy behavior is determined by the dominant component. However, the SOC induced coupling between the singlet and triplet components leads to a particle-hole asymmetric DOS. The disorder suppresses this particle hole asymmetry.

One of the key conclusions of this work is the  formation of odd frequency, spin triplet, even parity pairs in the normal metal segment of an SN junction with a triplet superconductor. Such odd frequency component only arises in the presence of SOC in the normal metal. The induced $\mathbf{d}$-vector of the odd frequency term is $\propto (\mathbf{\Sigma}_1 \times \mathbf{w})$, which is an even function of momentum. Proximity induced OTE superconductivity has been reported reported  for  $s$-wave superconductor junctions with topological insulators\cite{Yokoyama2012,Balatsky2012,Cayao2017} or with low dimensional  Rashba metals\cite{Reeg2015,Cayao2018}. In case of TI-superconductor junctions, either the gap modulation near the interface\cite{Balatsky2012,Cayao2017} or a finite exchange field\cite{Yokoyama2012} is essential for the emergence of OTE, and in the later case, the Andreev reflections give rise to OTE pairs which is a different mechanism. In contrast, the formation of  OTE pairs that we find for a triplet superconductor SN junction does not require  gap modulation near the interface or any exchange field.  We have obtained the full momentum-spin-energy structure of the OTE pairs. The OTE pairs have momentum dependence, which makes them vulnerable to impurity scattering.

OTE pairs have also been reported in the normal metal junctions with triplet superconductors where the normal metal did not have the SOC\cite{Tanaka2007}. These studies were performed using the Usadel equations with Nazarov-Tanaka boundary conditions.\cite{Nazarov1999,Tanaka2003} The present approach is different. We calculate the normal state Green function right at the interface, thus avoiding ambiguity with respect to possible boundary conditions. {The physical origin of the OTE pair formation differs in both approaches. In the quasi-classical approach, the underlying mechanism is Andreev reflection which leads to a mixing of parities at the interface. \cite{Reeg2015,Cayao2018} In contrast, the OTE pairs that we find in the tunneling matrix formalism are coming from a modification of triplet pairs in the two SOC generated helical bands.} 

In summary, we have investigated the effect of Rashba SOC on proximity induced superconductivity in the SN junctions consisting of conventional and unconventional superconductor.  Eq. \eqref{Eq:G11_trp_sbs} and \eqref{Eq:G12_trp_sbs} are very general, and the structure of the induced superconductivity is applicable to many other systems, such as surface states of  topological insulators or systems with Dresselhaus SOC. We examine the robustness of the induced superconductivity against disorder, and find that the induced triplet superconductivity gets suppressed by it. In contrast, the fully gapped $s$-wave superconductivity remains robust against disorder. We find that the OTE state is induced in the SN junctions with triplet superconductors, but it does not show any low energy signature. The OTE pairs may gets suppressed weakly or strongly by the disorder depending on their momentum structure.  We show that the formation of the OTE pairs requires a triplet superconductor in the SN junction, SOC and a favorable geometry. OTE pairs are  not induced in every triplet superconductor - normal metal  junction.

\begin{acknowledgments}
The authors are grateful to Shao-Pin Chiu and Juhn-Jong Lin for useful discussions. VM, YL and FCZ are partially supported by NSFC grants 11674278, 11920101005  and by the priority program of the Chinese Academy of Sciences grant No. XDB28000000, and  YL is also supported by the China Postdoctoral Science Foundation under grant No. 2020M670422 and by the Fundamental Research Funds for the Central Universities under grant No. E2E44305.  S.K. is supported by NSTC of Taiwan through Grant No. 112-2112-M-A49-MY4 and acknowledges support by the Yushan Fellowship Program of the Ministry of Education, Taiwan.
\end{acknowledgments}

\appendix
\section{General triplet  case}\label{App1}
This appendix provides the details of the normal metal Green's function 
for the general triplet case discussed in Sec.~\ref{sec:triplet}.
The $\hat{\mathbf{G}}_{11}$ component of the normal state Green's function  $\check{\mathbb{G}}$ is given by,
\begin{eqnarray}
\hat{\mathbf{G}}_{11} &= &\frac{b_+ b_-}{\mathbf{D}}  \left[ M_0 \sigma_0 - M_1 \mathbf{w}\cdot \boldsymbol{\sigma} - M_2 \mathbf{q}\cdot \boldsymbol{\sigma} \right. \nonumber \\
&& \left. - M_3 \Sigma_1 \tilde{\mathbf{d}}\cdot \boldsymbol{\sigma} - M_4  \Sigma_1 \tilde{\mathbf{d}}^\ast \cdot \boldsymbol{\sigma} \right]  \\ 
\mathbf{D} &=& M_0^2 - M_1^2 (\mathbf{w}\cdot \mathbf{w}) -M_2^2  (\mathbf{q}\cdot \mathbf{q})  \nonumber \\
&& - M_3^2 \Sigma_1^2 (\tilde{\mathbf{d}}\cdot \tilde{\mathbf{d}}) -M_4^2 \Sigma_1^2 (\tilde{\mathbf{d}}^\ast \cdot \tilde{\mathbf{d}}^\ast ) \nonumber \\
&& -2 M_1 M_2 \mathbf{q}\cdot \mathbf{w} - 4\epsilon_N M_1 \Sigma_1^2 \tilde{\mathbf{d}} \cdot \mathbf{w} \tilde{\mathbf{d}}^\ast \cdot \mathbf{w} \nonumber \\
& &- 2\epsilon_N^2 
\Sigma_1^4 | \tilde{\mathbf{d}} |^2 \tilde{\mathbf{d}} \cdot \mathbf{w} \tilde{\mathbf{d}}^\ast \cdot \mathbf{w}. 
\end{eqnarray}
Here $b_\pm = \bar{\omega}+\xi_{\mathbf{k}}\pm \tilde{\epsilon}_N$, $\mathbf{q}=i \tilde{\mathbf{d}}\times \tilde{\mathbf{d}}^\ast$, and  the coefficients $M_{i}$ ($i=0,..,4$) are,
\begin{eqnarray}
M_0 &=& a_0 b_+ b_- -b_0 \Sigma_1^2 |\tilde{\mathbf{d}} |^2 - \epsilon_N \mathbf{q} \cdot \mathbf{w}, \\
M_1 &=& -\epsilon_N b_+ b_- - \epsilon_N \Sigma_1^2 |\tilde{\mathbf{d}} |^2,  \\
M_2  &=& -b_0,   \\
M_3 &=& \epsilon_N \Sigma_1 \tilde{\mathbf{d}}^\ast \cdot \mathbf{w},  \\
M_4 &=& \epsilon_N \Sigma_1 \tilde{\mathbf{d}} \cdot \mathbf{w}. 
\end{eqnarray}
The anomalous component $\hat{\mathbf{G}}_{12}$ of $\check{\mathbb{G}}$ reads,
\begin{eqnarray}
\hat{\mathbf{G}}_{12} &= &  \hat{\mathtt{G}}_{12} \frac{i \sigma_y}{\mathbf{D}},
\end{eqnarray}
where 
\begin{eqnarray}
\hat{\mathtt{G}}_{12}  &=&\left[ C_0 \sigma_0 + C_1 \mathbf{w}\cdot \boldsymbol{\sigma} +C_2 {\Sigma}_1 \tilde{\mathbf{d}}  \cdot \boldsymbol{\sigma} \right.\nonumber \\
&& \left. + C_3 {\Sigma}_1 \tilde{\mathbf{d}} ^\ast \cdot \boldsymbol{\sigma} + C_4 \Sigma_1 \left( \tilde{\mathbf{d}}  \times \mathbf{w}\right) \cdot \boldsymbol{\sigma} \right],
\end{eqnarray}
\begin{eqnarray}
C_0 &=& 2\epsilon_N \xi b_+ b_- {\Sigma}_1 \tilde{\mathbf{d}}  \cdot \mathbf{w}  \nonumber \\
& & + 2\epsilon_N \Sigma_1^3 b_0 (\tilde{\mathbf{d}}  \times (\tilde{\mathbf{d}}  \times \tilde{\mathbf{d}} ^\ast)) \cdot \mathbf{w},\\
 C_1 &=& -2b_+ b_- \epsilon_N^2  {\Sigma}_1 \tilde{\mathbf{d}}  \cdot \mathbf{w} \nonumber \\
 & & - \epsilon_N^2 \Sigma_1^3 (\tilde{\mathbf{d}}  \times (\tilde{\mathbf{d}}  \times \tilde{\mathbf{d}} ^\ast)) \cdot \mathbf{w}, \\
 C_2 &=& b_+ b_- \left( a_0 b_0 + \epsilon_N^2 \right) \nonumber \\
 & & + \epsilon_N^2 \Sigma_1^2 \left( \tilde{\mathbf{d}}  \times \mathbf{w} \right)\cdot \left( \tilde{\mathbf{d}}^\ast \times \mathbf{w} \right), \\
 C_3 &=& -b_+ b_- \Sigma_1^2 \tilde{\mathbf{d}}  \cdot \tilde{\mathbf{d}}  \nonumber \\
 & & - \epsilon_N^2 \Sigma_1^2 \left( \tilde{\mathbf{d}} \times \mathbf{w} \right)\cdot \left( \tilde{\mathbf{d}}  \times \mathbf{w} \right), \\
 C_4 &=& -2i\epsilon_N \bar{\omega} b_+ b_- + i \epsilon_N^2 (\mathbf{q}\cdot \mathbf{w}).
\end{eqnarray}
Here $a_0=\bar{\omega}-\xi_{\mathbf{k}}$ and $b_0=\bar{\omega}+\xi_{\mathbf{k}}$. 

\section{Mixed parity}\label{App2}
Details for the normal and anomalous Green's functions the mixed parity state of Sec.~\ref{Sec:mx} are provided in this appendix.\\
The normal Green's function is,
\begin{eqnarray}
\hat{\mathbf{G}}_{11} &= & \frac{b_+ b_-}{\mathtt{D}} \left[ L_0 \sigma_0 - L_1 \widetilde{\mathbf{w}}\cdot \boldsymbol{\sigma} -L_2 \tilde{\mathbf{d}} \cdot \boldsymbol{\sigma}\right], \\
L_0 &=& a_0 b_+ b_- -b_0 \left(  \Sigma_s^2 + \Sigma_t^2 \tilde{\mathbf{d}} \cdot \tilde{\mathbf{d}}  \right) \nonumber \\
& & +2\tilde{\epsilon}_N \Sigma_s \Sigma_t \tilde{\mathbf{d}} \cdot \widetilde{\mathbf{w}}, \\
L_1 &=& -\tilde{\epsilon}_N \left( b_+ b_- - \Sigma_s^2 + \Sigma_t^2 \tilde{\mathbf{d}} \cdot \tilde{\mathbf{d}}   \right), \\
L_2 &=& -2\left( b_0 \Sigma_s \Sigma_t - \tilde{\epsilon}_N \Sigma_t^2 \tilde{\mathbf{d}} \cdot \widetilde{\mathbf{w}}   \right),\\
\mathtt{D}&=& L_0^2 -L_1^2 - L_2^2 \tilde{\mathbf{d}} \cdot \tilde{\mathbf{d}}  - 2L_1 L_2 \tilde{\mathbf{d}} \cdot \widetilde{\mathbf{w}} .
\end{eqnarray}
The anomalous Green's function can be cast into the form
\begin{eqnarray}
\hat{\mathbf{G}}_{12} &= & \hat{\mathtt{G}}_{12}  \frac{i \sigma_y}{\mathtt{D}},
\end{eqnarray}
with
\begin{eqnarray}
\hat{\mathtt{G}}_{12} &=& \left[ A_0 \sigma_0 + A_1 \widetilde{\mathbf{w}}\cdot \boldsymbol{\sigma} + A_2 \tilde{\mathbf{d}}\cdot \boldsymbol{\sigma} \right. \nonumber \\ 
&& \left. + A_3 ( \tilde{\mathbf{d}} \times \widetilde{\mathbf{w}}  )\cdot \boldsymbol{\sigma}   \right], \\
A_0 &=& (a_0 b_0-\tilde{\epsilon}_N^2) \Sigma_s - \Sigma_s^3 + \Sigma_s \Sigma_t^2 \tilde{\mathbf{d}}  \cdot \tilde{\mathbf{d}} \nonumber \\
&& +2\xi_{\mathbf{k}} \tilde{\epsilon}_N \tilde{\mathbf{d}} \cdot \widetilde{\mathbf{w}} \\
A_1  &=& 2\xi_{\mathbf{k}} \tilde{\epsilon}_N \Sigma_s -2\tilde{\epsilon}_N^2\Sigma_t \widetilde{\mathbf{w}} \cdot \tilde{\mathbf{d}} , \\
A_2 &=& \Sigma_t \left( a_0 b_0 +\tilde{\epsilon}_N^2 +\Sigma_s^2 - \Sigma_t^2 \tilde{\mathbf{d}} \cdot \tilde{\mathbf{d}}  \right)\ \\
A_3 &=&  -2i \tilde{\epsilon}_N \bar{\omega} \Sigma_t.
\end{eqnarray}
\section{Emergence of particle-hole asymmetry in the mixed parity SN junctions} \label{App3}
\begin{figure}[b]
\includegraphics[width=.85\linewidth]{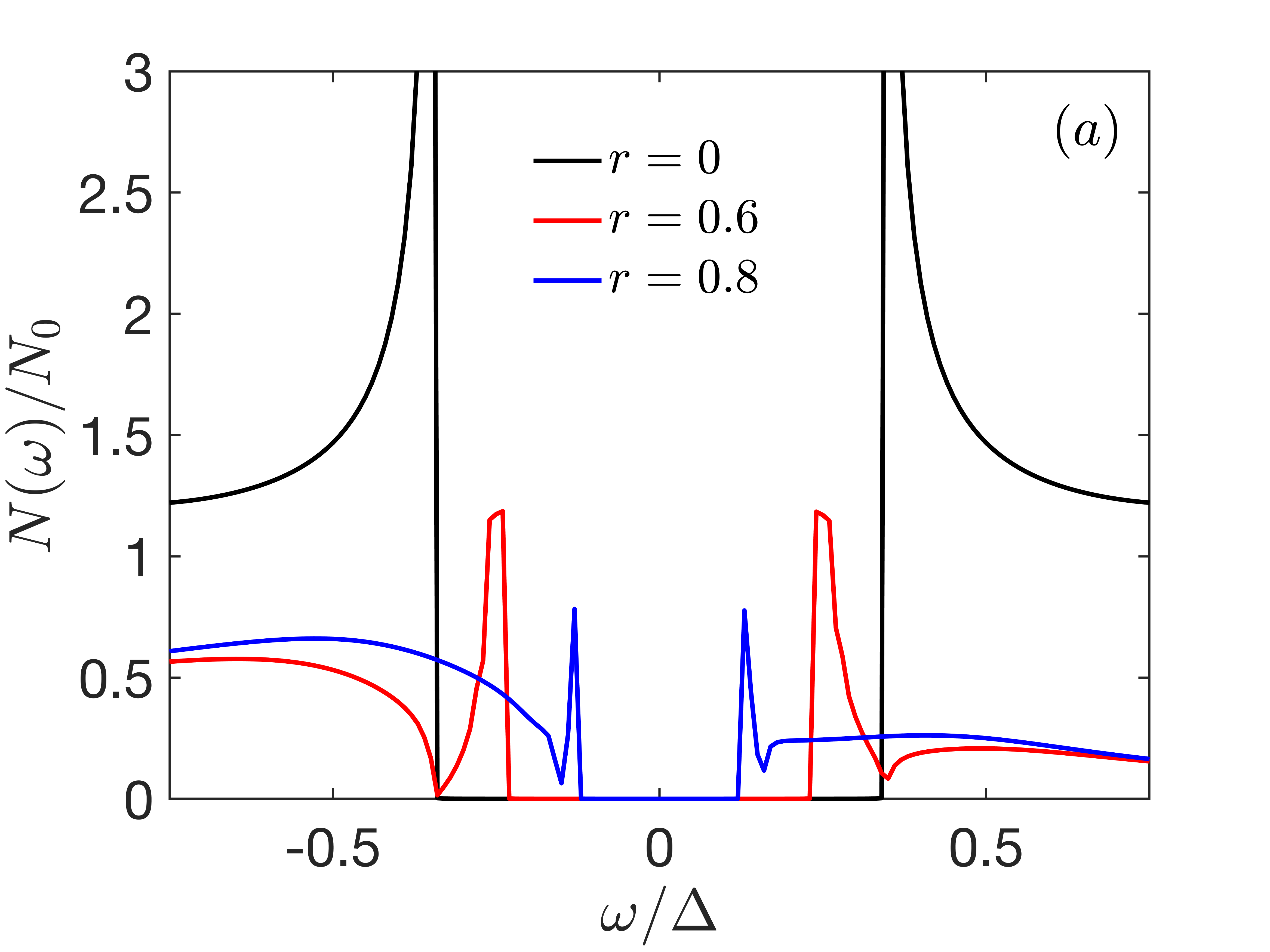}
\includegraphics[width=.85\linewidth]{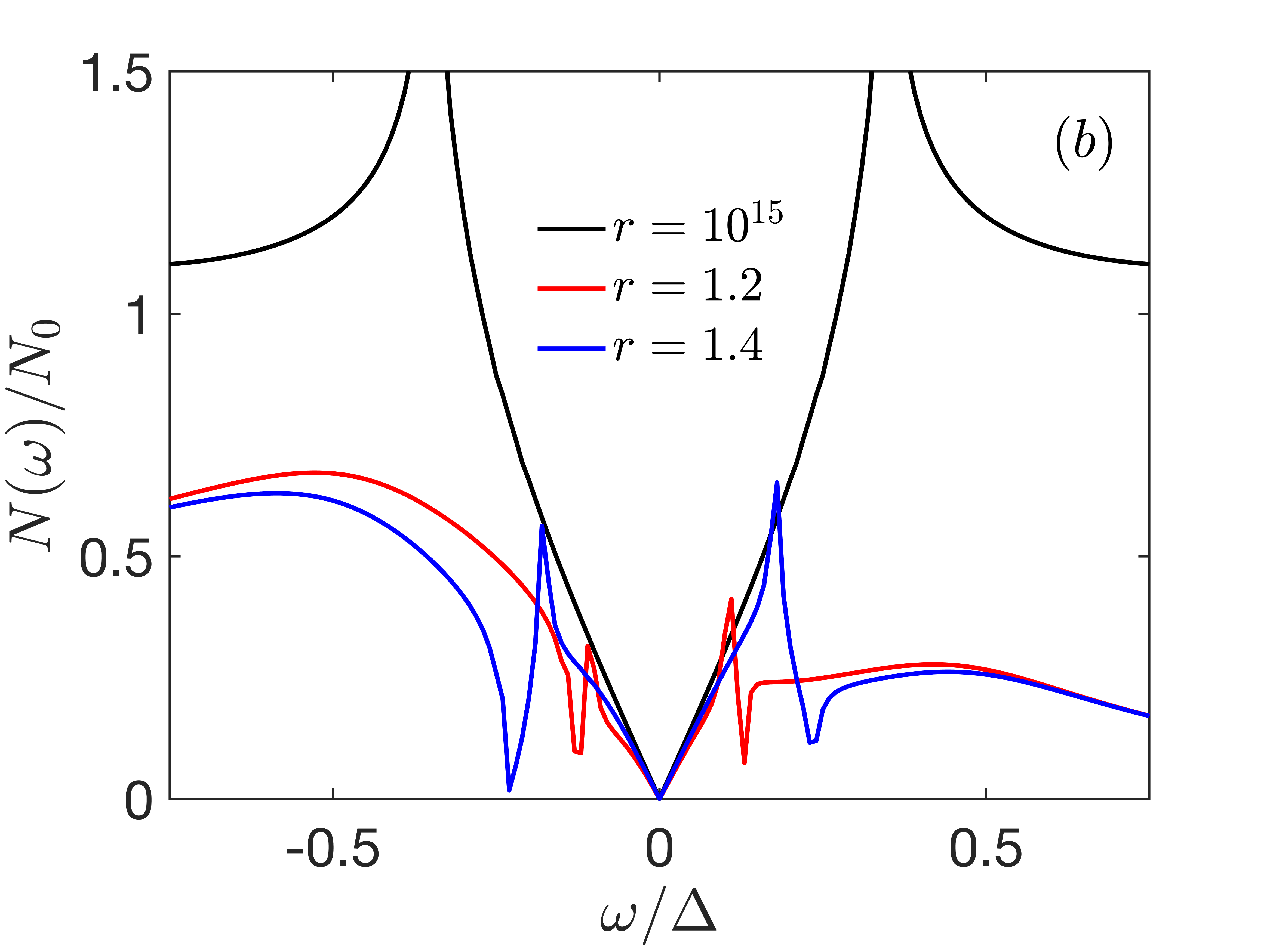}
\caption{The interface DOS for a mixed parity SN junction (a) singlet and (b) triplet case. The strong singlet $r=0$ in the panel (a) and strong $r=10^{15}$ in the panel (b) are particle hole symmetric, however, the near degenerate states show a particle-hole asymmetric DOS near the Fermi level. For these figures $\Gamma_t=0.5\Delta$ and $\epsilon_N=10\Delta$. }
\label{Fig:Mixed_ph_c}
\end{figure}

For a mixed parity superconductor, the limits of vanishing and infinitely large mixing parameter $r$, defined in Sec.\,\ref{Sec:mx}, recover the singlet and triplet cases respectively. This is explicitly demonstrated in Fig.\,\ref{Fig:Mixed_ph_c} where the interface density of states is shown for a wide range of $r$ values interpolating between $r=0$ and $r\rightarrow \infty$.
\clearpage
\bibliographystyle{apsrev4-1} 
\bibliography{SOC_PE_D}

\end{document}